\pdfoutput=1
\documentclass[manuscript,revised]{geophysics}
\usepackage{latexsym,amsmath,amssymb,amsthm}
\usepackage[toc]{appendix}
\renewcommand{\figdir}{Fig}

\usepackage{setspace}

\begin{document}

\title{Seismic Inversion by Newtonian Machine Learning}
\renewcommand{\thefootnote}{\fnsymbol{footnote}}

\ms{Seismic Inversion by Newtonian Machine Learning} 
\address{
\footnotemark[1] Department of Earth Science and Engineering, \\
King Abdullah University of Science and Technology, \\
Thuwal, Saudi Arabia, 23955-6900.
}
\author{Yuqing Chen\footnotemark[1] and Gerard T. Schuster\footnotemark[1]}

\footer{Example}
\lefthead{Chen \& Schuster}
\righthead{\emph{Seismic Inversion by Newtonian Machine Learning}}

\newpage
\begin{abstract}

We present a wave-equation inversion method that inverts skeletonized data for the subsurface velocity model. The skeletonized representation of the seismic traces consists of the low-rank latent-space variables predicted by a well-trained autoencoder neural network. The input to the autoencoder consist of the recorded common shot gathers, and the implicit function theorem is used to determine the perturbation of the skeletonized data with respect to the velocity perturbation. The gradient is computed by migrating the observed traces weighted by the residuals of the skeletonized data, and the final velocity model is the one that best predicts the observed latent-space parameters. We denote this hybrid inversion method as inversion by Newtonian machine learning because it inverts for the model parameters by combining the deterministic laws of Newtonian physics with the statistical capabilities of machine learning. Empirical results suggest that the cycle-skipping problem is largely mitigated compared to the conventional full waveform inversion (FWI) method by replacing the waveform differences by the those of the latent-space parameters. Numerical tests on both the synthetic and real data demonstrate the success of this skeletonized inversion method in recovering a low-wavenumber approximation to the subsurface velocity model. The advantage of this method over other skeletonized data methods is that no manual picking of important features is required because the skeletal data are automatically selected by the autoencoder. The disadvantage is that the inverted velocity model has less resolution compared to the FWI result, but which can be a good initial model for FWI. We suggest that the lowered resolution problem can be mitigated by using a multiscale method where the dimension of the latent space is gradually increased and more complexity is included into the input data. 

The most significant contribution of this paper is that it provides a general framework for using solutions to the governing PDE to invert skeletal data generated by any type of a neural network. The governing equation can be that for gravity, seismic waves, electromagnetic fields, and magnetic fields. The input data can be the records from different types of data and their skeletal features, as long as the model parameters are sensitive to their perturbations. The skeletal data can be the latent space variables of an autoencoder, a variational autoencoder, or a feature map from a convolutional neural network (CNN), or principal component analysis (PCA) features. In other words, we have combined the best features of Newtonian physics and the pattern matching capabilities of machine learning to invert seismic data by Newtonian machine learning.

\end{abstract}

\section{Introduction}

Full waveform inversion (FWI) has been shown to accurately invert seismic data for high-resolution velocity models \citep{lailly1983seismic, tarantola1984inversion, virieux2009overview}. However, the success of FWI heavily relies on a good initial model that is close to the true model, otherwise, cycle-skipping problems will trap the FWI in a local minimum\citep{bunks1995multiscale}. To mitigate the FWI cycle-skipping problem, \cite{bunks1995multiscale} proposed a multiscale inversion approach which initially inverts low-pass seismic data and then gradually admits higher frequencies as the iterations proceed. \cite{altheyab2015reflection} removed the mid- and far-offset cycle-skipped seismic traces before inversion and gradually incorporates them into the iteration solution as the velocity model become closer to the true model. \cite{wu2014seismic} use the envelope of the seismic traces to invert for the subsurface model as they claim that the envelope carries the ultra-low frequency information of the seismic data. \cite{ha2012laplace} invert the data in the Laplace-domain which is less sensitive to the lack of low frequencies than conventional FWI. \cite{sun1993time} and \cite{fu2017multiscale} use an amplitude replacement method to focus the inversion on reducing the phase mismatch instead of the waveform mismatch. In addition, they employ a multiscale approach by temporally integrating the traces to boost the low-frequencies and mitigate cycle-skipping problems, and then gradually introduce the higher frequencies as the iterations proceed.

The main reason non-linear inversion gets stuck in a local minimum is that the data are very complex (i.e, wiggly in time), which means that the objective function is very complex and characterized by many multiple minimums. To avoid this problem, \cite{luo1991wavea} suggested a skeletonized inversion method which combines the skeletonized representation of seismic data with the implicit function theorem to accelerate convergence to the vicinity of the global minimum \citep{lu2017tutorial}. Simplification of the data by skeletonization reduces the complexity of the misfit function and reduces the number of local minima. Examples of wave-equation inversion of skeletonized data include the following:

\begin{itemize}

\item \cite{lu2017tutorial} uses the solutions to the wave equation to invert the first-arrival traveltimes for the low-to-intermediate wavenumber details of the background velocity model. \cite{feng2019transmission} uses the traveltime misfit function to invert for both the subsurface velocity and anisotropic parameters in a vertical transverse isotropic medium.

\item Instead of minimizing the traveltime misfit function, \cite{li2016wave} finds the optimal S-velocity model that minimizes the difference between the observed and predicted dispersion curves associated with surface waves. \cite{liu20183d} extend 2D dispersion inversion of surface waves to the 3D case. 

\item Instead of inverting for the velocity model, \cite{dutta2016wave} developed a wave-equation inversion method that inverts for the subsurface $Q_{p}$ distribution. Here, they find the optimal $Q_{p}$ model by minimizing the misfit between the observed and the predicted peak/centroid-frequency shifts of the early arrivals. Similarly, \cite{li2017wave} utilize the peak frequency shift of the surface waves to invert for the $Q_{s}$ model.

\item A good tutorial for skeletonized inversion is by \cite{lu2017tutorial}.

\end{itemize}

One of the key problems with skeletonized inversion is that the skeletonized data must be picked from the original data, which can be labor intensive for large data sets. To overcome this problem, we propose obtaining the skeletonized data from an autoencoder, and then use solutions to the wave equation to invert such data for the model of interest \citep{schuster2018machine}. The skeletonized data correspond to the feature map in the latent space of the autoencoder, which has a reduced dimension and contains the significant parts of the input data related to the model. That is, we have combined the best features of Newtonian physics and the pattern matching capabilities of machine learning to invert seismic data by Newtonian machine learning. 

\begin{figure}[h]
\centering
\includegraphics[width=1\columnwidth]{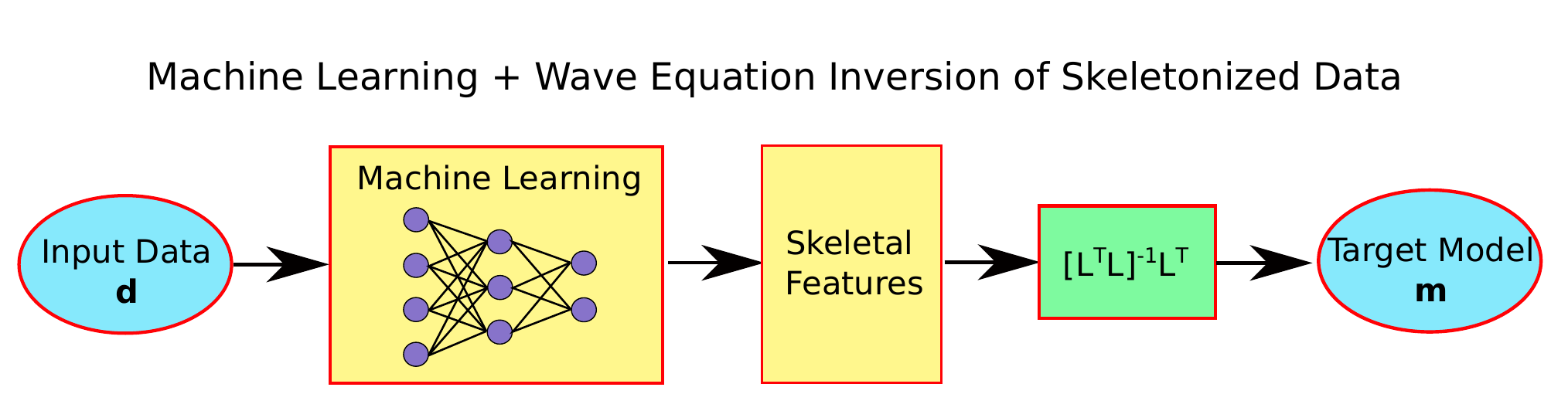}
\caption{The strategy for inverting the skeletonized latent variables.}
\label{fig:SKE_INV1}
\end{figure}

The autoencoder neural network is an unsupervised deep learning method that is trained for dimensionality reduction \citep{schmidhuber2015deep}. An autoencoder maps the data into a lower-dimensional space by extracting the data's most important features. It encodes the original data into a much more condensed representation, also denoted as the skeletonized representation, of the input data. The input data can be reconstructed by a decoder from the encoded value. In this paper, we first use the observed seismic traces as the training set to train the autoencoder neural network. Once the autoencoder is well trained, we feed both the observed and synthetic traces into the autoencoder to get their corresponding low-dimension representations. We build the misfit function as the sum of the squared differences between the observed and the predicted encoded value. To compute the gradient with respect to the model parameters such as the velocity in each pixel, we use the implicit function theorem to compute the perturbation of the skeletonized information with respect to the velocity. The high-level strategy for inverting the skeletonized latent variables is summarized in Figure \ref{fig:SKE_INV1}, where $\mathbf{L}$ corresponds to the forward modeling operator of the governing equations, such as the wave equation.

This paper is organized into four sections. After the introduction, we explain the theory of the wave equation inversion of seismic data skeletonized by an autoencoder. This theory includes the formulation first presented in \cite{luo1991wavea, luo1991waveb} where the implicit function theorem is used to employ numerical solutions to the wave equation for generating the Fr\'echet derivative of the skeletal data. We then present the numerical results for both synthetic data and field data recorded by a crosswell experiment. The last section provides a discussion, a summary of our work and its significance.  

\section{Theory}
Conventional full waveform inversion (FWI) inverts for the subsurface distribution by minimizing the $l^{2}$ norm of the waveform difference between the observed and synthetic data. However, this misfit function is highly nonlinear and the iterative solution often gets stuck into the local minima \citep{bunks1995multiscale}. To mitigate the problem, skeletonized inversion methods simplify the objective function by combining the skeletonized representation of data, such as the traveltimes, with the implicit function theorem, to give a gradient optimization method that quickly converges to the vicinity of the glocal minimum. Instead of manually picking the skeletonized data, we allow the unsupervised autoencoder to generate it.

\subsection{Theory of Autoencoder} 

\begin{figure}[h]
\centering
\includegraphics[width=1\columnwidth]{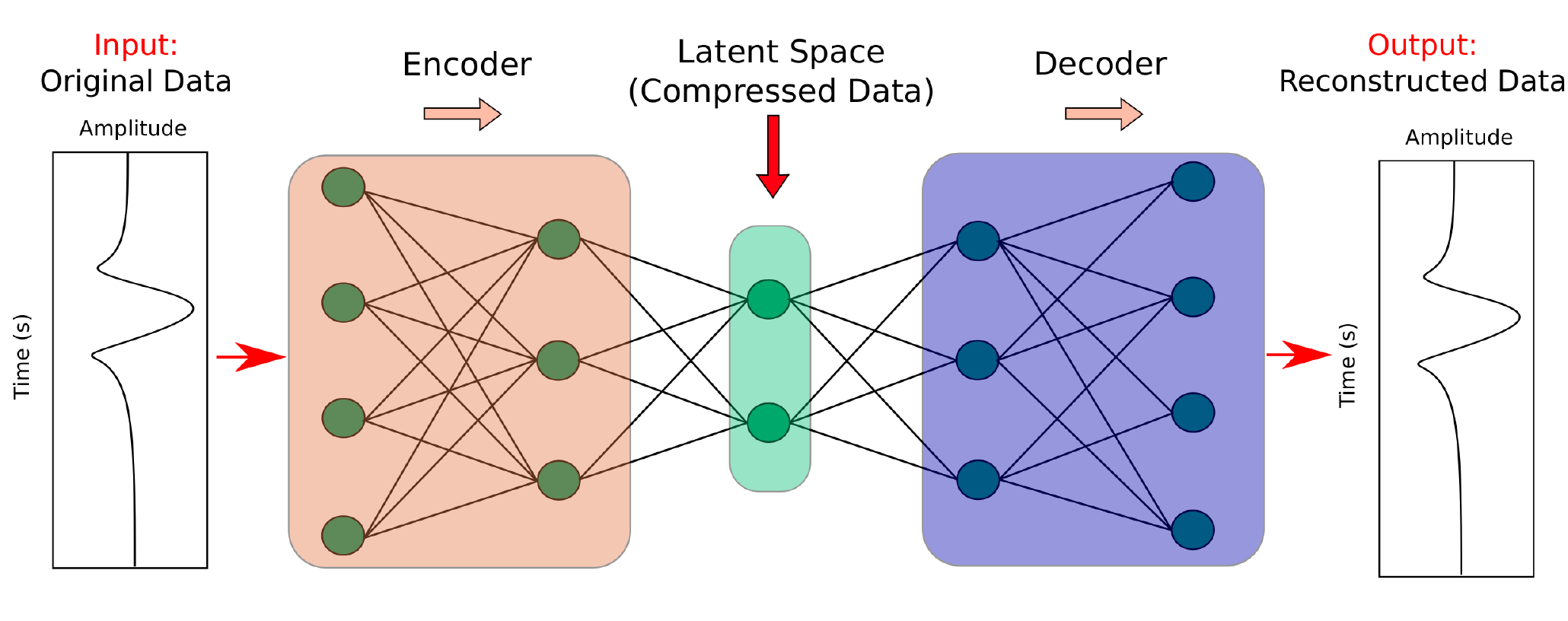}
\caption{An example of an autoencoder architecture with two layers for encoder and two layer for decoder. The dimension of the latent space is two.}
\label{fig:AE1}
\end{figure}

An autoencoder is an unsupervised neural network in which the predicted output is the same as the input data, as illustrated in Figure \ref{fig:AE1}. An autoencoder is trained to learn the extremely low-dimensional representation of the input data, also denoted as the skeletonized representation, in an unsupervised manner. It is similar to the principal component analysis (PCA), which is generally used to represent input data using a smaller dimensional space than originally present \citep{hotelling1933analysis}. However, PCA is restricted to finding the optimal rotation of the original data axes that maximizes its projections to the principal components axes. In comparison, the autoencoder with a sufficient number of layers can find almost any non-linear sparse mapping between the input and output images. A typical autoencoder architecture is shown in Figure \ref{fig:AE1} which generally includes three parts: the encoder, the latent space, and the decoder.  

\begin{itemize}
\item Encoder: Unsupervised learning by an autoencoder uses a set of training data consisting of N training samples $\{\mathbf{x}^{(1)},\mathbf{x}^{(2)},..., \mathbf{x}^{(N)}\}$, where $\mathbf{x}^{(i)}$ is the $i^{th}$ feature vector with dimension $D\times 1$ and $D$ represent the number of features for each feature vector. The encoder neural network indicated by the pink box in Figure \ref{fig:AE1} encodes the high-dimension input data $\mathbf{x}^{(i)}$ into a low-dimension latent space with dimension $C \times 1$ using a series of neural layers with a decreasing number of neurons; here $C$ is smaller than $D$. This encoding operations can be mathematically described as $\mathbf{z}^{(i)}=g\big( \mathbf{W}_{1}\mathbf{x}^{(i)}+\mathbf{b}_{1}\big)$ , where $\mathbf{W}_{1}$ and $\mathbf{b}_{1}$ represent the model parameter and the vector of bias terms for the first layer, and $g()$ indicates the activation function such as a sigmoid, ReLU, Tanh and so on.


\item Latent Space: The compressed data $\mathbf{z}^{(i)}$ with dimension $C \times 1$ in the latent space layer (emphasized by the green box) is the lowest dimension space in which the input data is reduced and the key information is preserved. The latent space usually has a few neurons which forces the autoencoder neural network to create effective low-dimension representations of the high-dimension input data. These low-dimension attributes can be used by the decoder to reconstruct the original input.

\item Decoder: The decoder portion of the neural network represented by the purple box reconstructs the input data from the latent space representation $\mathbf{z}^{(i)}$ by a series of neural network layers with an increasing number of neurons. The reconstructed data $\mathbf{\tilde{x}^{(i)}}$ are calculated by $\mathbf{\tilde{x}}^{(i)}=\mathbf{W}_{2} \mathbf{z}^{(i)}+\mathbf{b}_{2}$, where $\mathbf{W}_{2}$ and $\mathbf{b}_{2}$ represent the model parameter and the bias term of the decoder neural network, respectively.
\end{itemize}

The parameters of the autoencoder neural network are determined by finding the values of $\mathbf{w}_{i}$ and $\mathbf{b}_{i}$ for $i=1,2$ that minimize the following objective function: 
\begin{align}
J(\mathbf{W}_{1},\mathbf{b}_{1},\mathbf{W}_{2},\mathbf{b}_{2}) &= \sum_{i=1}^{N}(\mathbf{\tilde{x}}^{(i)} -\mathbf{x}^{(i)})^{2}, \\ \nonumber
&= \sum_{i=1}^{N} \bigg(\mathbf{W}_{2}\big(g(\mathbf{W}_{1}\mathbf{x}^{(i)}+\mathbf{b}_{1})\big)+\mathbf{b}_{2}-\mathbf{x}^{(i)} \bigg)^{2}.
\label{eq1}
\end{align}
In practice a preconditioned steepest descent method is used for mini-batch inputs.

\subsection{Skeletonized Representation of Seismic Data by Autoencoder}

To get the low-dimension skeletonized representation of seismic data by the autoencoder, the input data consist of seismic traces, each with the dimension of $nt \times 1$. In this case, each seismic trace is defined as one training example in the training set generated by a crosswell seismic experiment. For the crosswell experiment, there are $N_{s}$ sources in the source well and $N_{r}$ receivers in the receiver well. We mainly focus on the inversion of the transmitted arrivals by windowing the input data around the early arrivals.  

\begin{figure}[h]
\centering
\includegraphics[width=0.8\columnwidth]{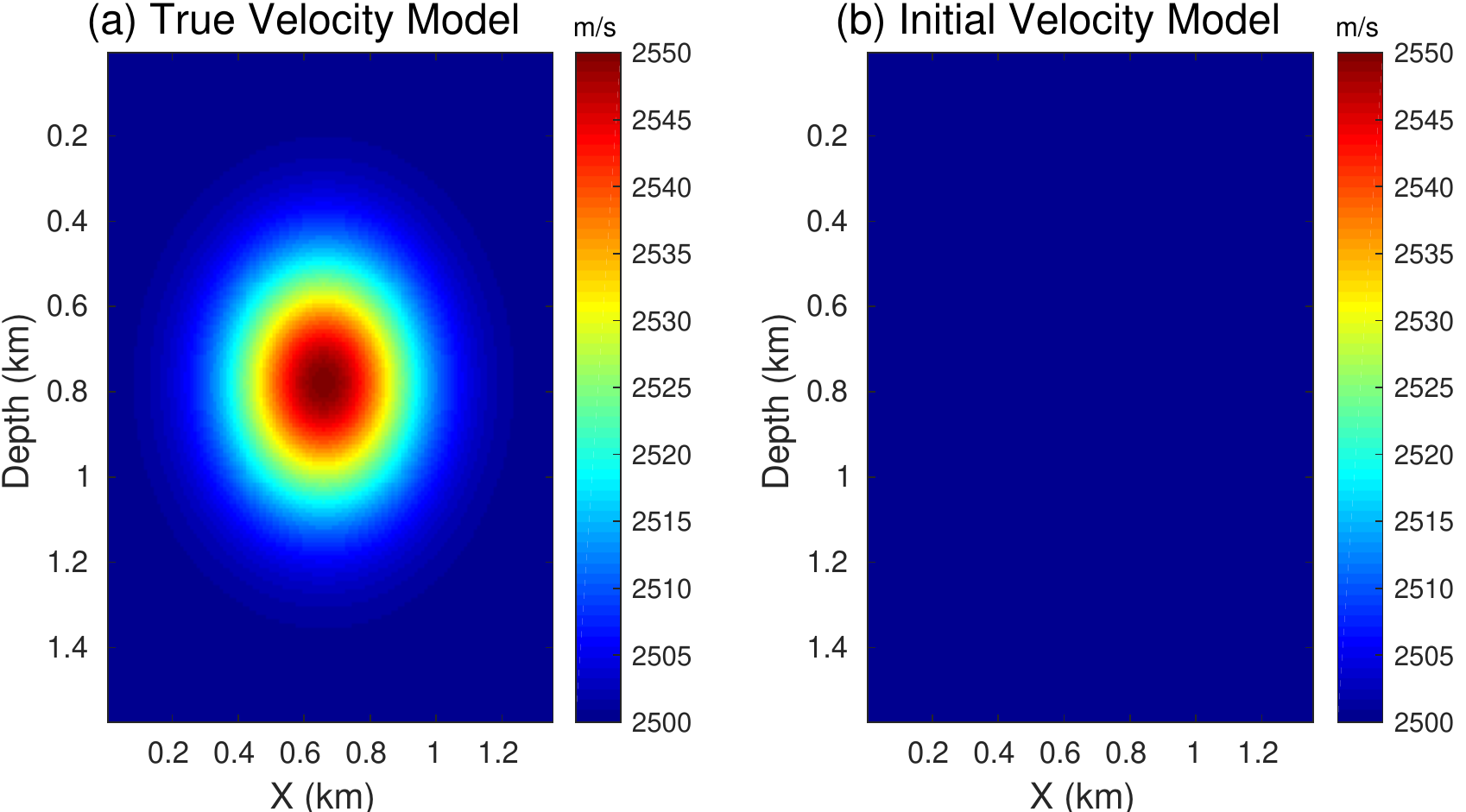}
\caption{A homogeneous velocity model with a Gausian velocity anormaly in the center.}
\label{fig:AE2}
\end{figure}

Figure \ref{fig:AE2}a shows a homogeneous velocity model with a Gaussian anomaly in the center. Figure \ref{fig:AE2}b is the corresponding initial velocity model which has the same background velocity as the true velocity model. A crosswell acquisition system with two 1570-m-deep cased wells separated by 1350 m is used as the source and receiver wells. The finite-difference method is used to compute 77 acoustic shot gathers for both the observed and synthetic data with 20 m shot intervals. Each shot is recorded with 156 receivers that are evenly distributed along the depth at a spacing of 10 m. To train the autoencoder network, we use the following workflow.

\begin{figure}[h]
\centering
\includegraphics[width=0.8\columnwidth]{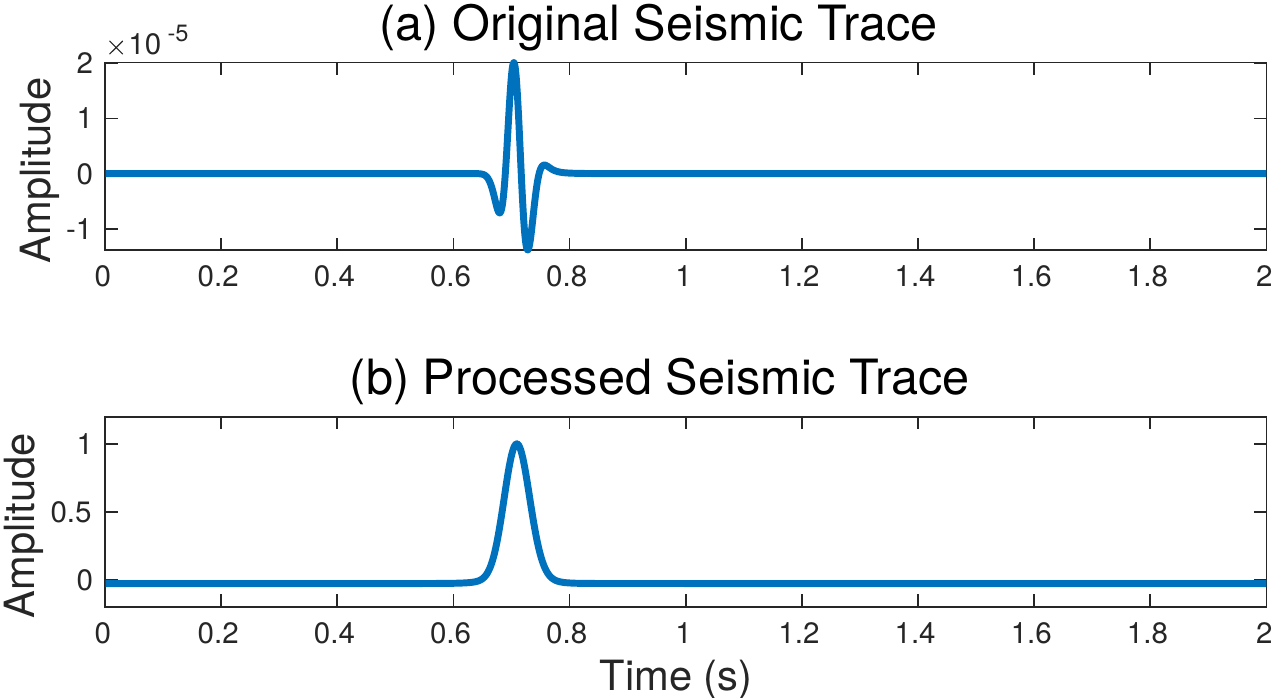}
\caption{The (a) orginal seismic trace and the (b) processed seismic trace.}
\label{fig:Pro1}
\end{figure}

\begin{figure}[h]
\centering
\includegraphics[width=0.8\columnwidth]{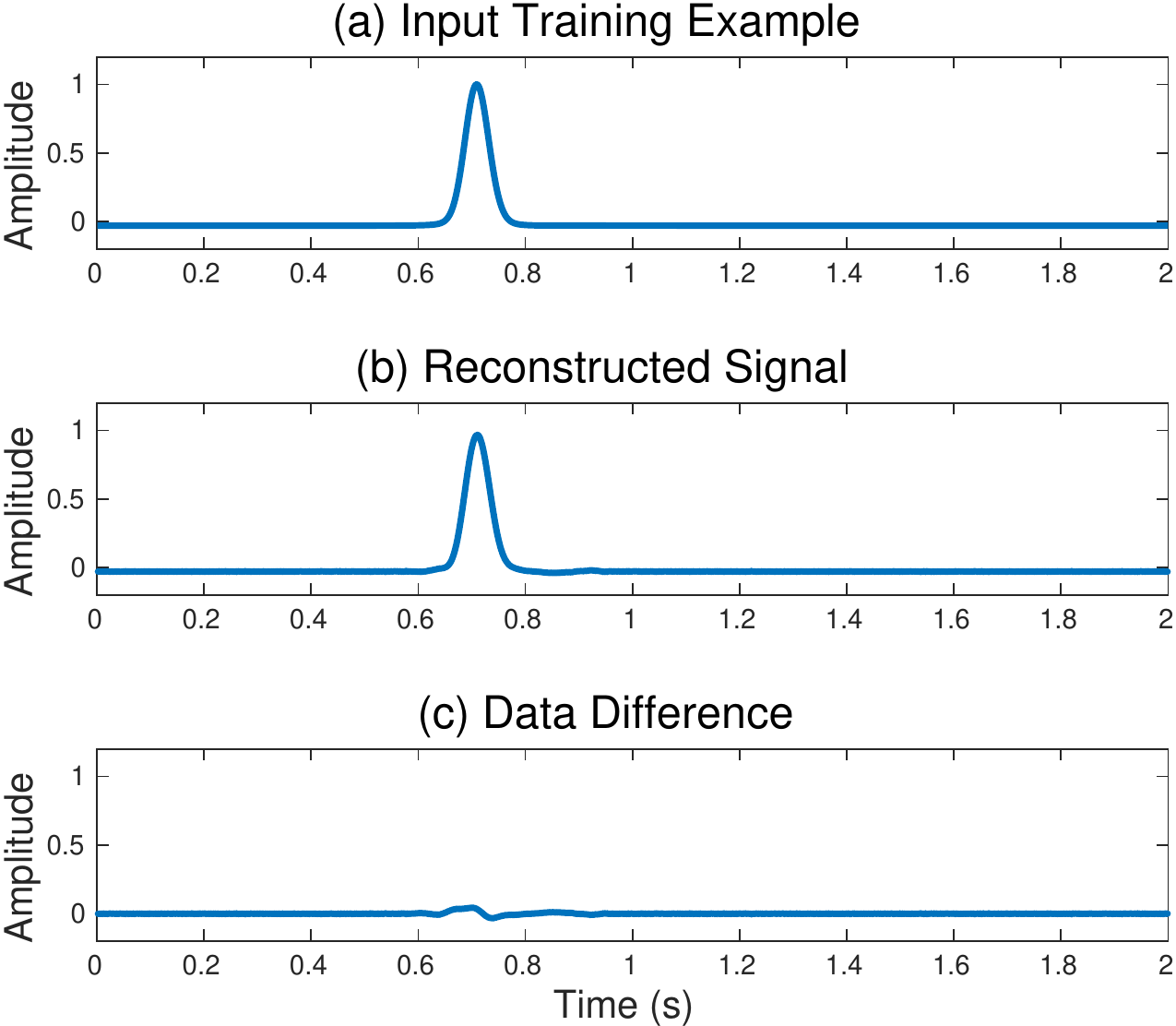}
\caption{The (a) input training example, (b) reconstructed signal by autoencoder and their difference.}
\label{fig:Pro2}
\end{figure}

\begin{enumerate}

\item Build the training set. For every five observed shots, we randomly select one shot gather as part of the training set that consist of a total of 2496 training examples, or seismic traces. We didn't use all the shot gathers for training because of the increase in the computation cost.

\item Data processing. Each seismic trace is Hilbert transformed to get its envelope then subtracted by their mean and divided by their variance. Figure \ref{fig:Pro1}a and \ref{fig:Pro1}b show a seismic trace before and after processing, respectively. We use the signal envelope instead of the original seismic trace because it is less complicated than the original signal. And according to our tests, the signal envelope leads to faster convergence compared to the original seismic signal. 

\item Training the autoencoder. We feed the processed training set into an autoencoder network where the dimension of its latent space is equal to 1. In other words, each training example with a dimension of $nt \times 1$ will be encoded as a smaller number of latent variables by the encoder. The autoencoder parameters are updated by iteratively minimizing equation 1. The Adam and mini-batch gradient descent methods are used to train this network. Figure \ref{fig:Pro2}a and \ref{fig:Pro2}b show an input training example and its corresponding reconstructed signal by the autoencoder, respectively, and their difference is shown in Figure \ref{fig:Pro2}c. 

\end{enumerate}

\begin{figure}[h]
\centering
\includegraphics[width=0.95\columnwidth]{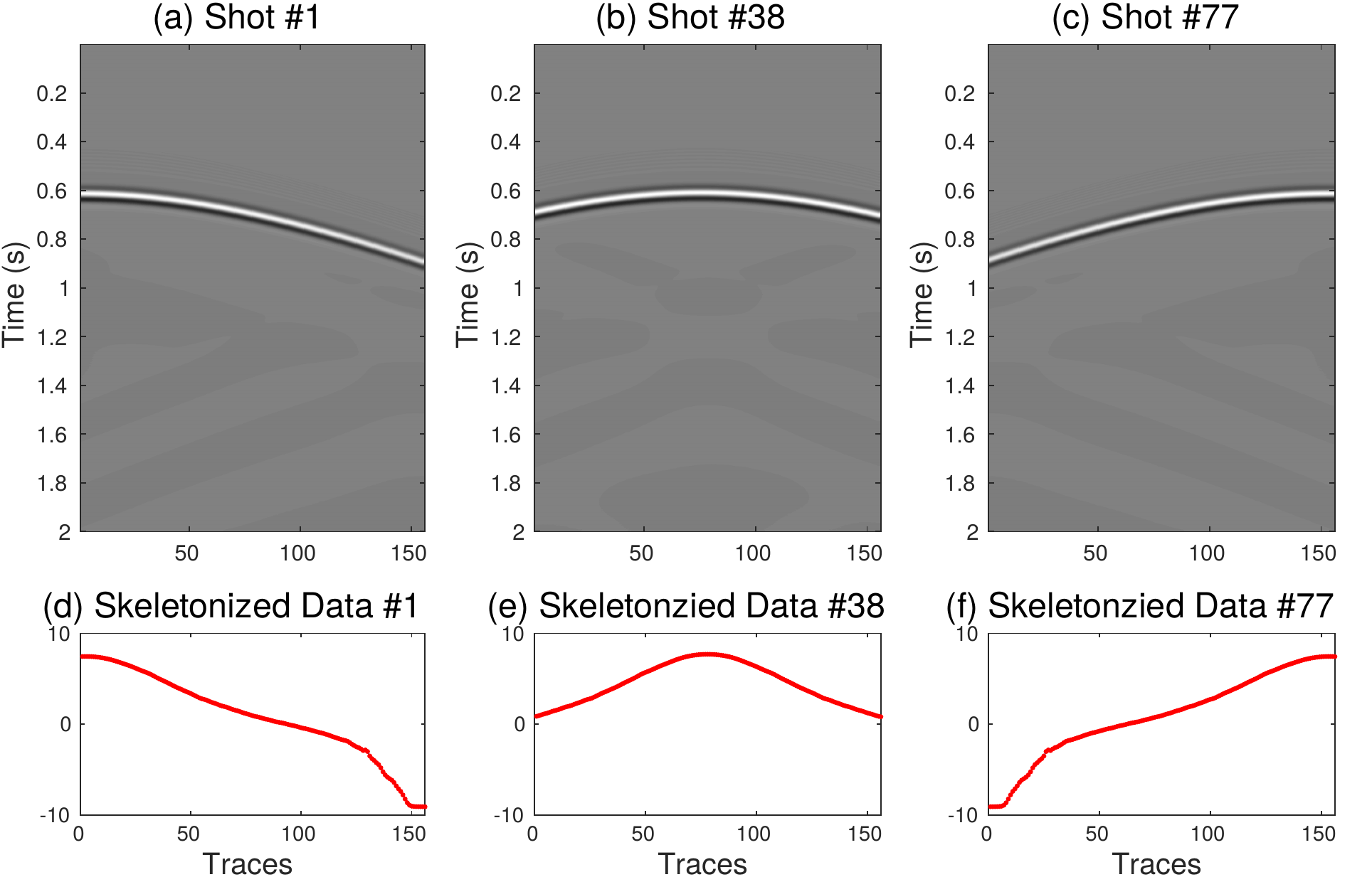}
\caption{Three shot gathers with their corresponding encoded data.}
\label{fig:AE3}
\end{figure}

After training is finished, we input all the observed and predicted seismic traces into the well-trained autoencoder network to get their skeletonized low-dimensional representation. Of course, each input seismic trace requires the same data processing procedure as we did for the training set. Figure \ref{fig:AE3}a, \ref{fig:AE3}b and \ref{fig:AE3}c shows three observed shot gathers which are not included in the training set, and their encoded values are shown in Figure \ref{fig:AE3}d, \ref{fig:AE3}e and \ref{fig:AE3}f which are the skeletonized representations of the input seismic traces. The encoded values do not have any units and can be considered as a skeletonized attribute of the data. However, the autoencoder believes that these encoded values are the best low-dimension representation of the original input in the least-square sense. 

\begin{figure}[h]
\centering
\includegraphics[width=1.1\columnwidth]{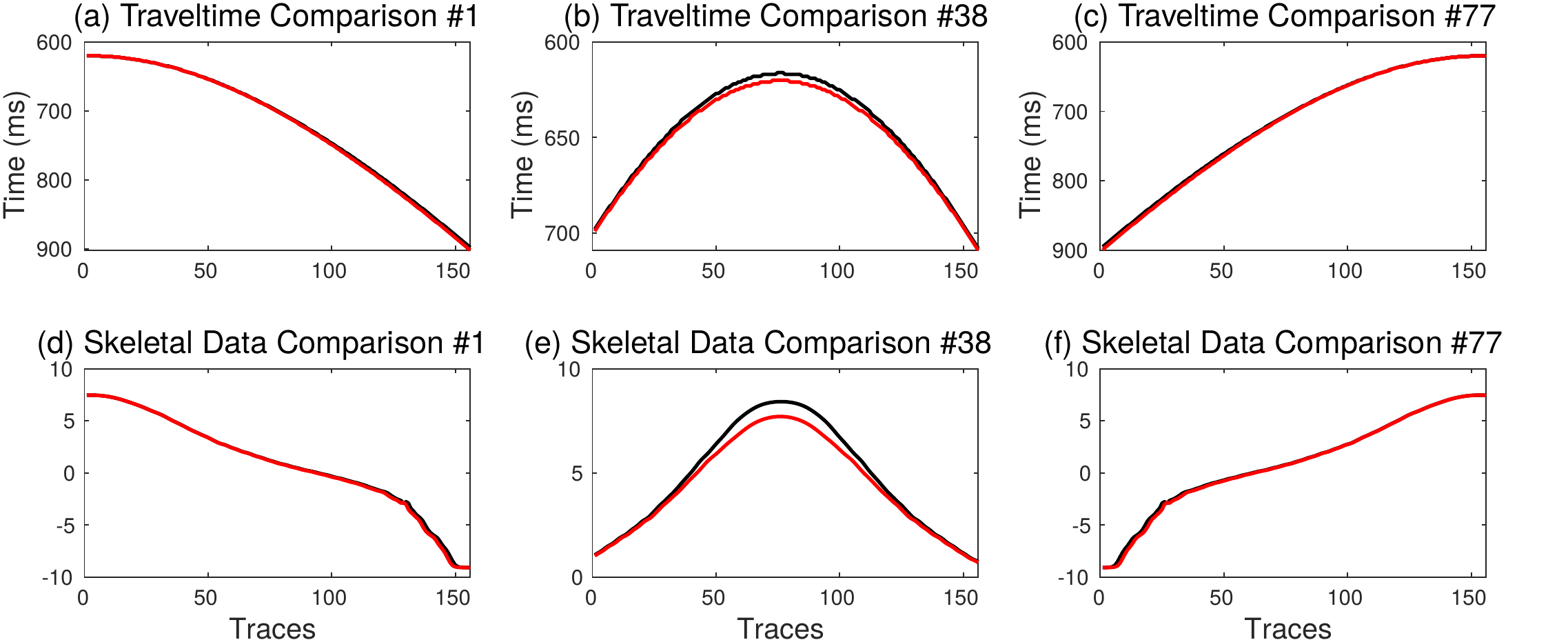}
\caption{The comparison of the traveltime misfit functions and skeletal data misfit functions for different shot gathers. The black and red curves represent the observed and synthetic data, respectively.}
\label{fig:Pro3}
\end{figure}

We compare the traveltime differences and the encoded low-dimension representation differences for the observed and synthetic data in Figure \ref{fig:Pro3}. The black and red curves represent the observed and synthetic data, respectively. Figure \ref{fig:Pro3}b shows a larger traveltime difference than Figure \ref{fig:Pro3}a and \ref{fig:Pro3}c as its propagating waves are affected more by the Gaussian anomaly than the other two shots. However, the misfit function for the low-dimensional representation of the seismic data exhibits a pattern similar to that of the traveltime misfit function. Both reveal a large misfit at the traces affected by the velocity anomaly. Similar to the traveltime misfit values, the encoded values are also sensitive to the velocity changes. In this case, we can conclude that the (1) autoencoder network is able to estimate the effective low-dimension representation of the input data and (2) the encoded low-dimensional representation can be used as a skeletonized feature sensitive to changes in the velocity model.

\subsection{Theory of the Skeletonized Inversion with Autoencoder}
In order to invert for the velocity model from the skeletonized data, we use the implicit function theorem to compute the perturbation of the skeletonized data with respect to the velocity. 

\subsubsection{Connective Function}
A cross-correlation function is defined as the connective function that connects the skeletonized data with the pressure field. This connective function measures the similarity between the observed and synthetic traces as 

\begin{equation}
f_{z_{1}}(\mathbf{x}_{r},t;\mathbf{x}_{s}) = \int dt p_{z-z_{1}}(\mathbf{x}_{r},t;\mathbf{x}_{s})_{obs} p_{z}(\mathbf{x}_{r},t;\mathbf{x}_{s})_{syn},
\label{eq2}
\end{equation}
where $p_{z}(\mathbf{x}_{r},t;\mathbf{x}_{s})_{syn}$ represents a synthetic trace for a given background velocity model recorded at the receiver location $\mathbf{x}_{r}$ due to a source excited at location $\mathbf{x}_{s}$. The subscript $z$ is the skeletonized feature (low-dimension representaion of the seismic trace) that is encoded by a well-trained autoencoder network. Similarly, $p_{z-z_{1}}(\mathbf{x}_{r},t;\mathbf{x}_{s})_{obs}$ denotes the observed trace with a encoded skeletonized feature equal to $z-z_{1}$ that has the same source and receiver location as $p_{z}(\mathbf{x}_{r},t;\mathbf{x}_{s})_{syn}$, and $z_{1}$ is the distance between the synthetic and observed skeletal data in the latent space.

For an accurate velocity model, the observed and synthetic traces will have the same encoded values in the latent space. Therefore, we seek to minimize the distance in the latent space between an observed and synthetic traces. This can be done by finding the shift value $z_{1}=\Delta z$ that maximizes the crosscorrelation function in equation \ref{eq2}. If $\Delta z = 0$, it indicates that the correct velocity model has been found and the synthetic and observed traces have the same encoded values in the latent space. The $\Delta z$ that maximizes the crosscorrelation function in equation \ref{eq2} should satisfy the condition that the derivative of $f_{z_{1}}(\mathbf{x}_{r},t;\mathbf{x}_{s})$ with respect to $z_{1}$ is equal to zero.  Thus,

\begin{align}
\dot{f}_{\Delta z} &= \bigg[\frac{\partial f_{z_{1}(\mathbf{x}_{r},t;\mathbf{x}_{s})}}{\partial z_{1}}\bigg]_{z_{1}=\Delta z}, \\ \nonumber
& = \int dt \dot{p}_{z-\Delta z}(\mathbf{x}_{r},t;\mathbf{x}_{s})_{obs}p_{z}(\mathbf{x}_{r},t;\mathbf{x}_{s})_{syn}=0,
\label{eq3}
\end{align}
where $\dot{p}_{z-\Delta z}(\mathbf{x}_{r},t;\mathbf{x}_{s})_{obs} = \partial p_{z-z_{1}}(\mathbf{x}_{r},t;\mathbf{x}_{s})/\partial z_{1}$. Equation 3 is the connective function that acts as an intermediate equation to connect the seismogram with the skeletonized data, which are the encoded values of the seismograms \citep{luo1991wavea, luo1991waveb}. Such a connective function is necessary because there is no wave equation that relates the skeletonized data to a single type of model parameters \citep{dutta2016wave}. The connective function will be later used to derive the derivative of skeletonized data with respect to the velocity.

\subsubsection{Misfit Function}

\begin{figure}[h]
\centering
\includegraphics[width=0.6\columnwidth]{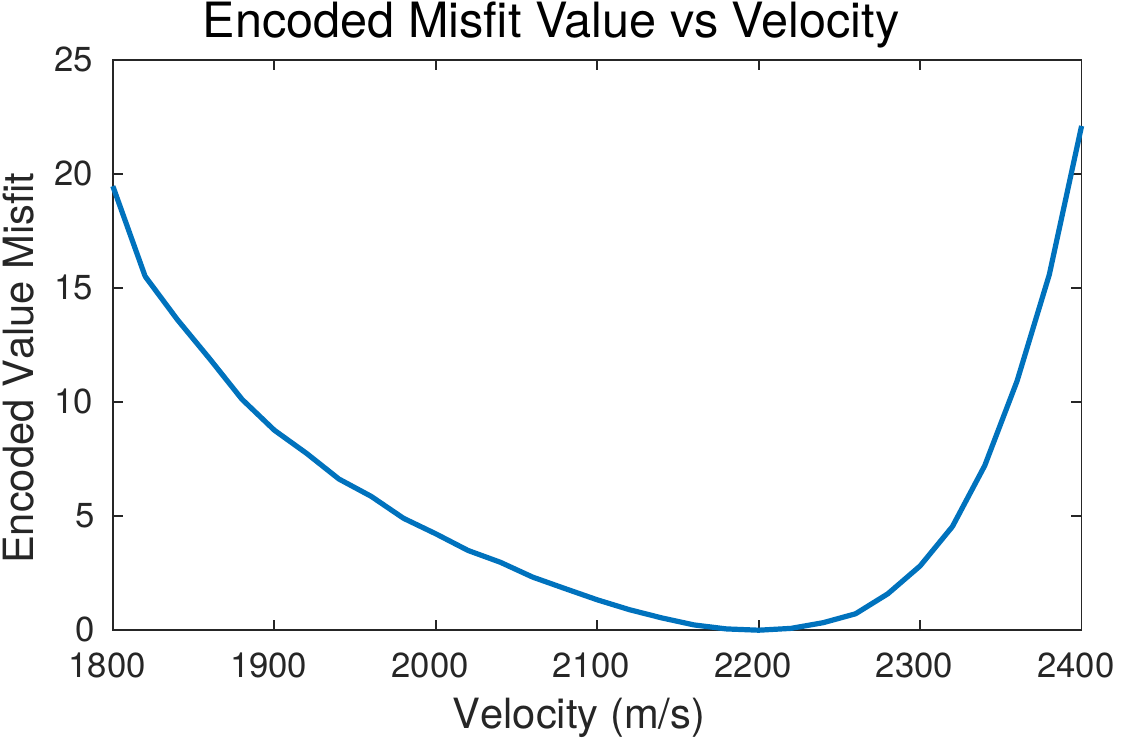}
\caption{Plot of the encoded value misfit function versus hypothetical velocity values for the velocity model. The observed data is generated with $v=2200\  m/s$.}
\label{fig:Misfit1}
\end{figure}

The misfit function of the skeletonized inversion with the autoencoder method is defined as

\begin{equation}
\epsilon = \frac{1}{2} \sum_{s} \sum_{r} \Delta z(\mathbf{x}_{r},\mathbf{x}_{s})^{2},
\label{eq4}
\end{equation}
where $\Delta z$ is the difference of the encoded value in the latent space between the observed and synthetic data. The gradient $\gamma (\mathbf{x})$ is given by

\begin{equation}
\gamma(\mathbf{x}) = -\frac{\partial \epsilon}{\partial v(\mathbf{x})}=-\sum_{s}\sum_{r}\frac{\partial \Delta z}{\partial v(\mathbf{x})}\Delta z(\mathbf{x}_{r},\mathbf{x}_{s}).
\label{eq5}
\end{equation}
Figure \ref{fig:Misfit1} shows the encoded value misfit versus different values of velocity, which clearly shows that the misfit monotonically decreases as the velocity value approaches to the correct velocity value ($v = 2200\ m/s$). Therefore, the skeletonized misfit function in equation \ref{eq5} is able to quickly converge to the global minimum when using the gradient optimization method. Using equation 3 and the implicit function theorem we can get

\begin{align}
\frac{\partial \Delta z}{\partial v(\mathbf{x})}&=\frac{\big[\frac{\partial \dot{f}_{\Delta z}}{\partial v(\mathbf{x})}\big]}{\big[\frac{\partial \dot{f}_{\Delta z}}{\partial \Delta z}\big]}, \\ \nonumber
&= \frac{1}{E}\int dt \dot{p}_{z-\Delta z}(\mathbf{x}_{r},t;\mathbf{x}_{s})_{obs}\frac{\partial p_{z}(\mathbf{x}_{r},t;\mathbf{x}_{s})_{syn}}{\partial v(\mathbf{x})},
\label{eq6}
\end{align}
where
\begin{equation}
E =\int dt \ddot{p}_{z-\Delta z}(\mathbf{x}_{r},t;\mathbf{x}_{s})_{obs}p_{z}(\mathbf{x}_{r},t;\mathbf{x}_{s})_{syn}.
\label{eq7}
\end{equation}
The Fr\'echet derivative $\partial p_{z}(\mathbf{x}_{r},t;\mathbf{x}_{s})/\partial v(\mathbf{x})$ is derived in the next section.

\subsubsection{Fr\'echet Derivative}
The first-order acoustic wave-equation can be written as

\begin{align}
&\frac{\partial p}{\partial t} + \rho c^{2} \nabla \cdot \mathbf{v} = S(\mathbf{x}_{s},t), \\ \nonumber
&\frac{1}{\rho}\nabla p + \frac{\partial \mathbf{v}}{\partial t}=0,
\label{eq8}
\end{align}
where $p$ represents the pressure, $\mathbf{v}$ represents the particle velocity, and $\rho$ and $c$ indicate the density and velocity, respectively. $S(\mathbf{x}_{s},t)$ denotes a source excited at location $\mathbf{x}_{s}$ and at the excitation time 0 and the listening time is $t$. To derive the formula for the Fr\'echet derivative of the pressure field with respect to the perturbation in velocity $c(\mathbf{x})$, we linearize the wave equation in equation 8. A perturbation of $c(\mathbf{x}) \rightarrow c(\mathbf{x}) + \delta c(\mathbf{x})$ will produce a perturbation in pressure $p(\mathbf{x}) \rightarrow p(\mathbf{x})+\delta p(\mathbf{x})$ and particle velocity $\mathbf{v}(\mathbf{x}) \rightarrow \mathbf{v}(\mathbf{x})+\delta \mathbf{v}(\mathbf{x})$, which satisfy the linearized acoustic equation given by

\begin{align}
&\frac{\partial \delta p}{\partial t} + \rho c^{2} \nabla \cdot \delta \mathbf{v} = - 2\rho c \delta c \nabla \cdot \mathbf{v}, \\ \nonumber
&\frac{1}{\rho}\nabla \delta p + \frac{\partial \delta \mathbf{v}}{\partial t}=0.
\label{eq9}
\end{align}
Using the Green's function $g_{p}(\mathbf{x}_{r},t;\mathbf{x},0)$, the solution of equation 9 can be written as 

\begin{equation}
\delta p(\mathbf{x}_{r},t;\mathbf{x}_{s})=-\big(2\rho c g_{p}(\mathbf{x}_{r},t;\mathbf{x},0)*\nabla \cdot \mathbf{v}(\mathbf{x},t;\mathbf{x}_{s}))\delta c(\mathbf{x}\big), 
\label{eq10}
\end{equation} 
where $*$ indicates convolution operator in time. Dividing by $\delta c(\mathbf{x})$ on both sides, we get
\begin{equation}
\frac{\delta p(\mathbf{x}_{r},t;\mathbf{x}_{s})}{\delta c(\mathbf{x}\big) } =-2\rho c g_{p}(\mathbf{x}_{r},t;\mathbf{x},0)*\nabla \cdot \mathbf{v}(\mathbf{x},t;\mathbf{x}_{s}).
\label{eq11}
\end{equation}
Substituting equation \ref{eq11} into equation 6 we get

\begin{equation}
\frac{\partial \Delta z}{\partial c(\mathbf{x})} = -\frac{1}{E} \int dt (2\rho c g_{p}(\mathbf{x}_{r},t;\mathbf{x},0)*\nabla \cdot \mathbf{v}(\mathbf{x},t;\mathbf{x}_{s})) \times \dot{p}_{z-\Delta z}(\mathbf{x}_{r},t;\mathbf{x}_{s})_{obs}.
\label{eq12}
\end{equation}
Substituting equation \ref{eq12} into equation \ref{eq5}, the gradient of $\gamma (\mathbf{x})$ can be expressed as

\begin{align}
\gamma(\mathbf{x}) &= -\sum_{s}\sum_{r}\frac{\partial \Delta z}{\partial c (\mathbf{x})}\Delta z(\mathbf{x}_{r},\mathbf{x}_{s}), \\ \nonumber
&=\sum_{s}\sum_{r}\frac{1}{E}\int dt (2\rho c g_{p}(\mathbf{x}_{r},t;\mathbf{x},0)*\nabla \cdot \mathbf{v}(\mathbf{x},t;\mathbf{x}_{s}))\times \dot{p}_{z-\Delta z}(\mathbf{x}_{r},t;\mathbf{x}_{s})_{obs}\Delta z(\mathbf{x}_{r},\mathbf{x}_{s}), \\ \nonumber
&=\sum_{s}\sum_{r} \frac{1}{E}\int dt (2\rho c g_{p}(\mathbf{x}_{r},t;\mathbf{x},0)*\nabla \cdot \mathbf{v}(\mathbf{x},t;\mathbf{x}_{s}))\times \Delta p_{z}(\mathbf{x}_{r},t;\mathbf{x}_{s}),
\label{eq13}
\end{align}
where $\Delta p_{z}(\mathbf{x}_{r},t;\mathbf{x}_{s}) = \dot{p}_{z-\Delta z}(\mathbf{x}_{r},t;\mathbf{x}_{s})_{obs}\Delta z(\mathbf{x}_{r},\mathbf{x}_{s})$ denotes the data residual which is obtained by weighting the derivative of the observed trace with respect to the latent variable $z$. Then the difference of observed and predicted encoded values $\Delta z$ are scaled by a factor of $E$. Using the identity

\begin{equation}
\int dt [f(t)*g(t)]h(t) = \int dt g(t) [f(-t)*h(t)], 
\label{eq14}
\end{equation}
equation 13 can be rewritten as 
\begin{align}
\gamma(\mathbf{x}) &= -2\rho c \sum_{s}\sum_{r}\int dt \nabla \cdot \mathbf{v}(\mathbf{x},t;\mathbf{x}_{s}) \big(g_{p}(\mathbf{x}_{r},-t;\mathbf{x},0)*\Delta p_{z}(\mathbf{x}_{r},t;\mathbf{x}_{s})\big), \\ \nonumber
&=-2 \rho c \sum_{s}\int dt \nabla \cdot \mathbf{v}(\mathbf{x},t;\mathbf{x}_{s}) \sum_{r} \big(g_{p}(\mathbf{x}_{r},-t;\mathbf{x},0)*\Delta p_{z}(\mathbf{x}_{r},t;\mathbf{x}_{s})\big), \\ \nonumber
&=-2\rho c \sum_{s} \int dt \nabla \cdot \mathbf{v}(\mathbf{x},t;\mathbf{x}_{s})q(\mathbf{x},t;\mathbf{x}_{s}),
\label{eq15}
\end{align}
where q is the adjoint-state variables of $p$ \citep{plessix2006review}. Equation 15 is the gradient of the skeletonized data which can be numerically calculated by a zero-lag crosscorrelation of a forward-wavefield $\nabla \cdot \mathbf{v}(\mathbf{x},t;\mathbf{x}_{s})$ with the backward-propagated wavefield $q(\mathbf{x},t;\mathbf{x}_{s})$. The velocity model is updated by the steepest gradient descent method

\begin{equation}
c(\mathbf{x})_{k+1} = c(\mathbf{x})_{k} + \alpha_{k} \gamma(\mathbf{x})_{k},
\label{16}
\end{equation}
where $k$ indicates the iteration number and $\alpha_{k}$ represents the step length.
\section{NUMERICAL TEST}
The effectiveness of wave equation inversion of data skeletonized by the autoencoder method is now demonstrated with two synthetic data and with crosswell data collected by Exxon in Texas \citep{chen1990subsurface}. The synthetic data are generated for a crosswell acquisition system using a 2-8 finite-difference solutions to the acoustic wave equation. 

\subsection{Crosswell Layer Model}

\begin{figure}[h]
\centering
\includegraphics[width=0.9\columnwidth]{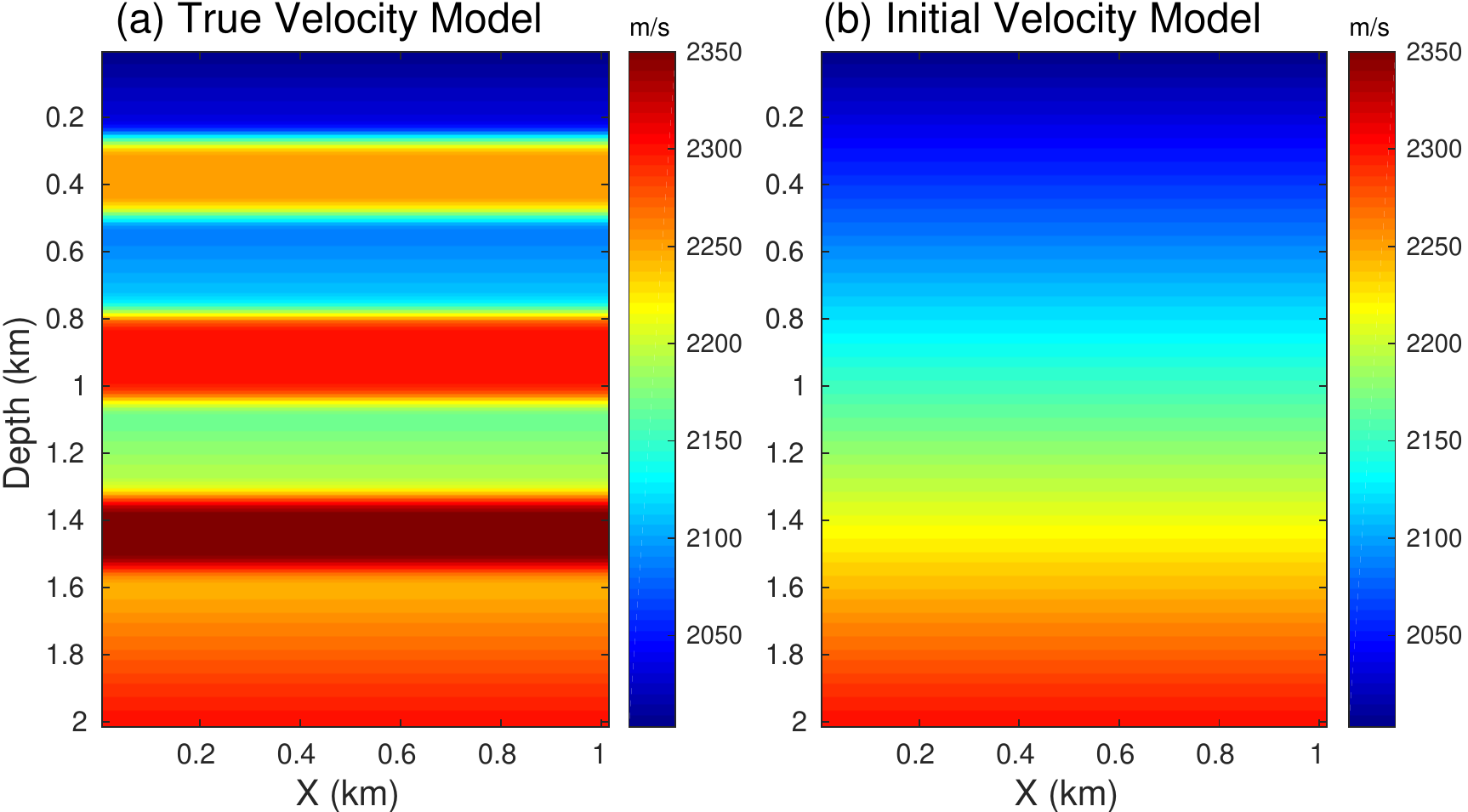}
\caption{The (a) true velocity and (b) linear increasing initial models.}
\label{fig:Layer1}
\end{figure}

Skeletonized inversion with the autoencoder is now tested on a layered model and a crosswell acquisition geometry. Figure \ref{fig:Layer1}a shows the true velocity model which has three high-velocity horizontal layers and a linear increasing background velocity. A Ricker wavelet with a peak frequency of 15 Hz is used as the source wavelet. A fixed-spread crosswell acquisition geometry is deployed where 99 shots at a source interval of 20 m are evenly distributed along a vertical well located at x = 10. The data are recorded by 200 receivers for each shot, where the receivers are uniformly distributed every 10 m in depth along with a receiver well located 1000 m away from the source well. The simulation time of the seismic data is 2 s with a time interval of 1 ms. 

\begin{figure}[h]
\centering
\includegraphics[width=1\columnwidth]{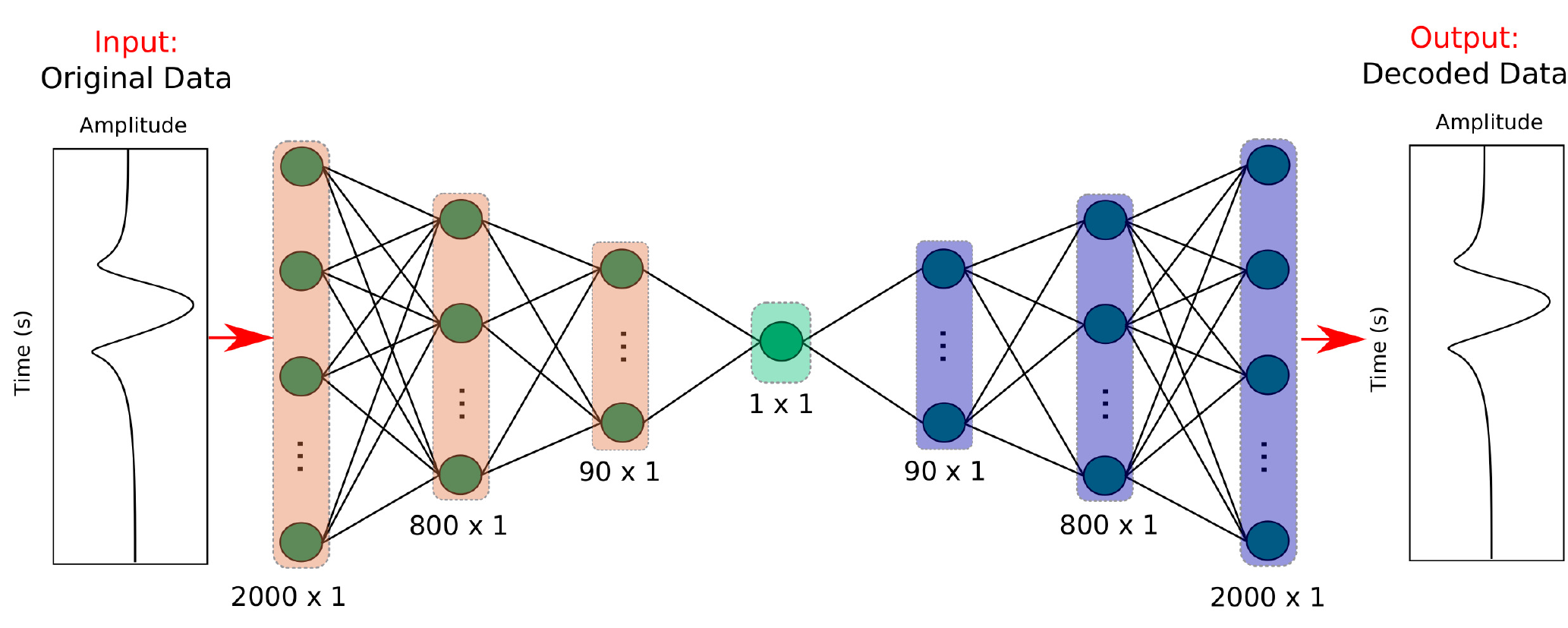}
\caption{The architecture of the autoencoder neural network.}
\label{fig:Layer4}
\end{figure}

The training set includes 4000 observed seismic traces because every five shots we take one shot gather as part of the training data. After data processing, we feed the training data into the autoencoder network shown in Figure \ref{fig:Layer4}. The number below each layer indicates the dimension of that layer. The boxes with pink, green and blue colors represent the encoder network, latent space and decoder network, respectively. The autoencoder network is trained with mini-batches of 50 traces. We use Tanh activation function instead of ReLU because the input data have both positive and negative parts. The whole training progress only takes several minutes on a workstation with 56 cores and 1 GPU. 

\begin{figure}[h]
\centering
\includegraphics[width=0.9\columnwidth]{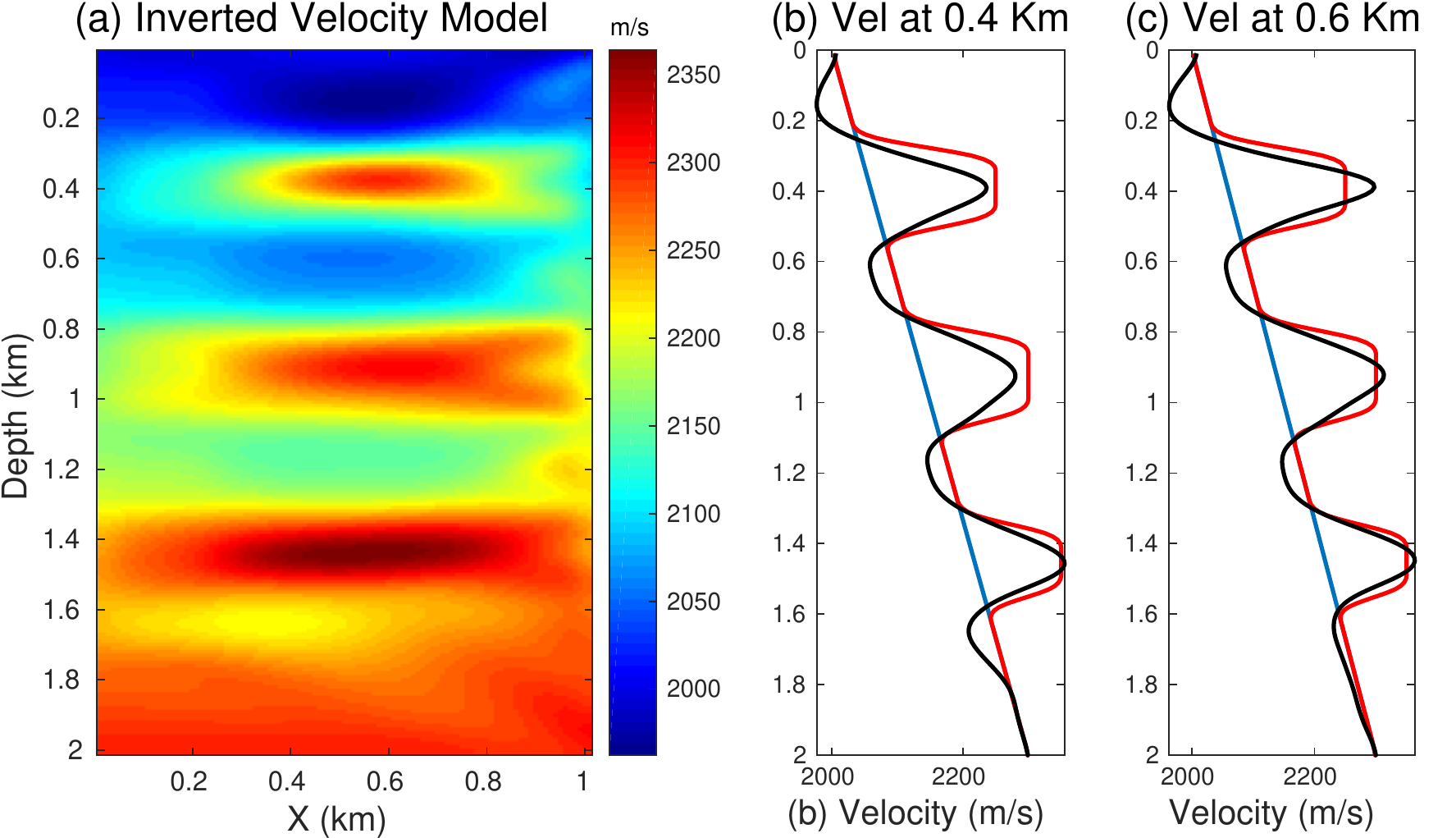}
\caption{The (a) inverted velocity model and the comparison of the vertical velocity profiles at (b) x = 0.4 km and x = 0.6 km. The blue, red and black curve indicate the velocity profiles of the initial, true and inverted velocity model, respectively.}
\label{fig:Layer2}
\end{figure}

\begin{figure}[h]
\centering
\includegraphics[width=0.7\columnwidth]{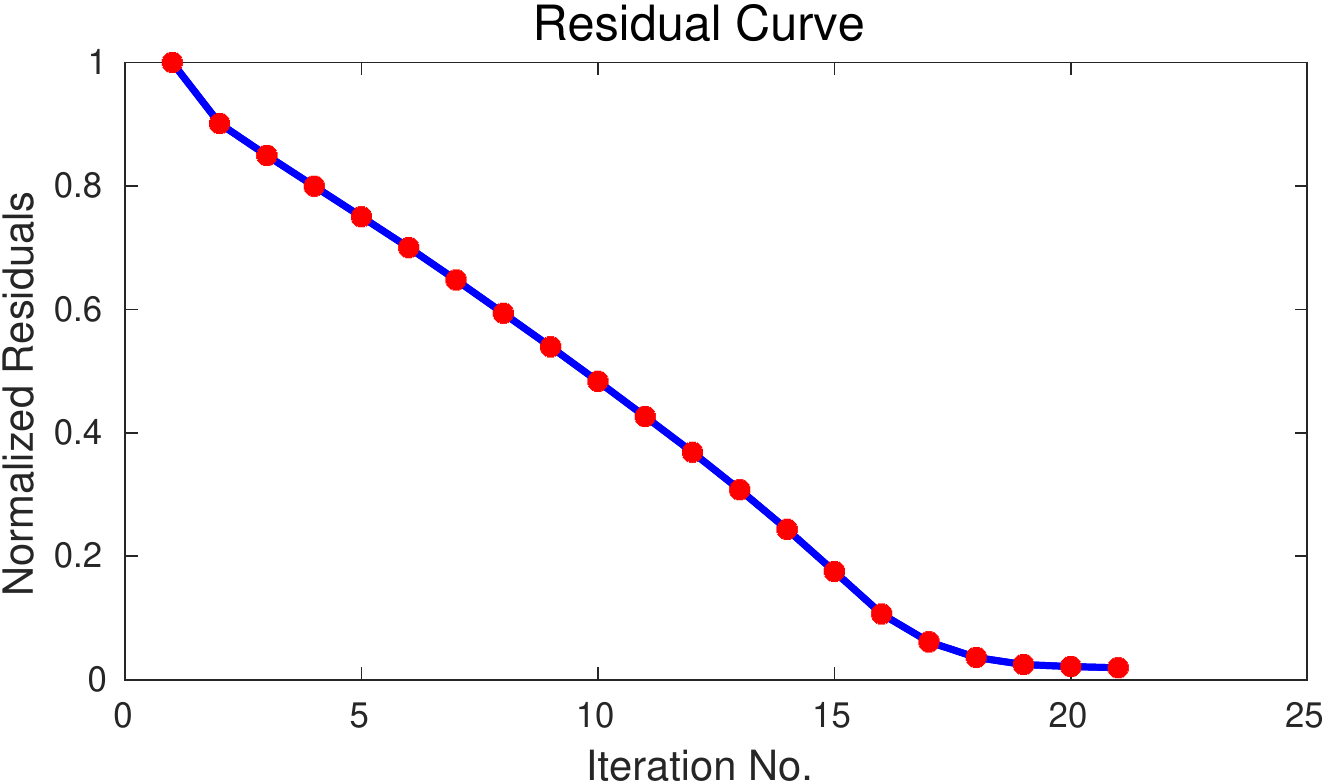}
\caption{The normalized data residual versus iteration numbers.}
\label{fig:Layer3}
\end{figure}

After the autoencoder neural network is well trained, we can simply input the synthetic traces generated at each iteration of the inversion to get their encoded values. Therefore the skeletonized misfit and gradient functions can be calculated in order to update the velocity model. Figure \ref{fig:Layer1}b shows a linear increasing initial model and the inverted result is in Figure \ref{fig:Layer2}a, which successfully recovers the three high-velocity layers. To further check the correctness of the inverted result, we compared the vertical velocity profiles between the initial, true and inverted velocity model at x = 0.4 km and x = 0.6 km. The blue, red and black lines in Figures \ref{fig:Layer2}b and \ref{fig:Layer3}c represent the velocity profiles from the initial, true and inverted velocity models, respectively. Figure \ref{fig:Layer3} shows the normalized data residual plotted against the iteration number, which clearly shows a fast convergence to the global minimum.

\subsection{Crosswell Marmousi Model}

\begin{figure}[h]
\centering
\includegraphics[width=0.92\columnwidth]{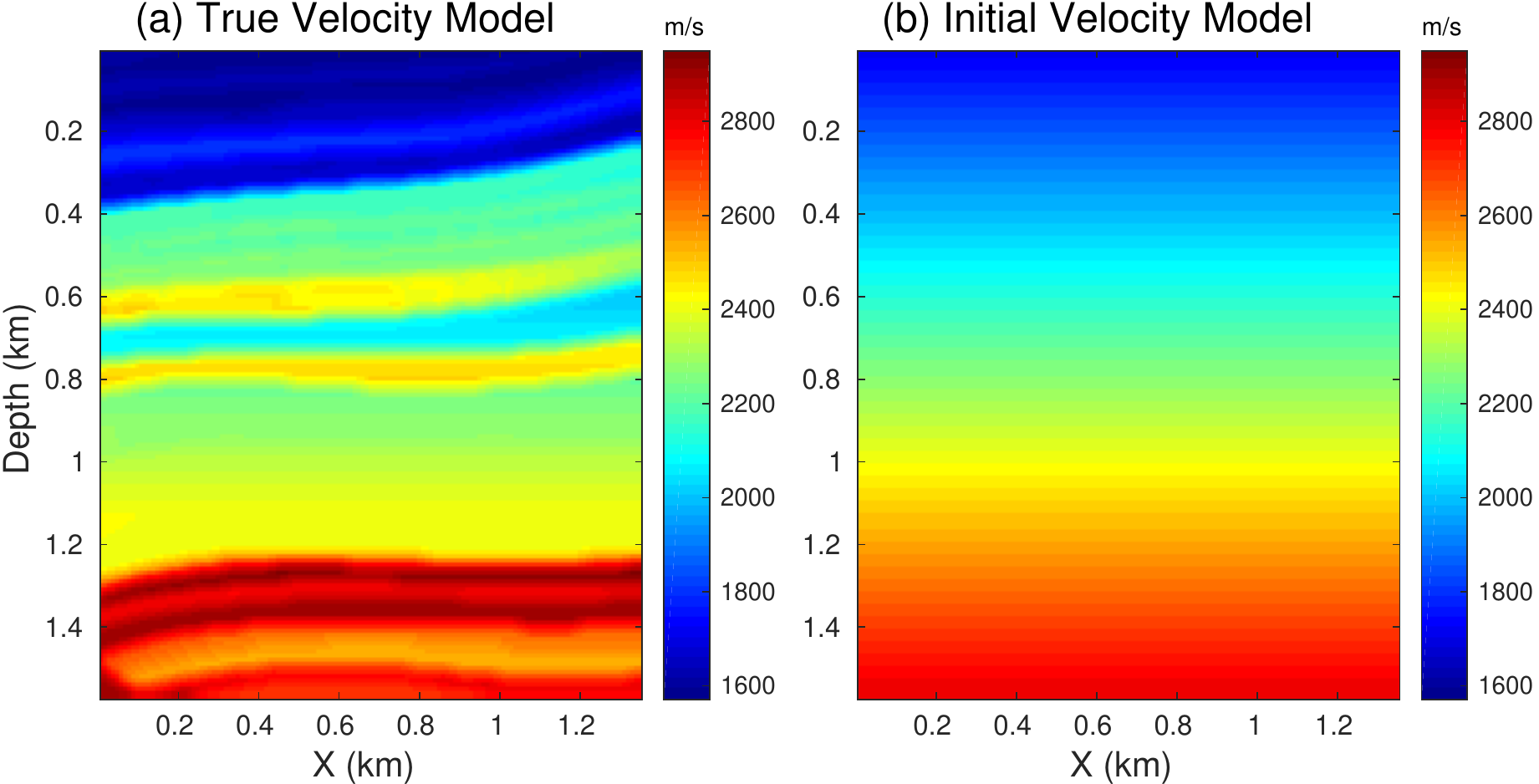}
\caption{The (a) true velocity model and (b) linear increasing initial model.}
\label{fig:Mar1}
\end{figure}

\begin{figure}[h]
\centering
\includegraphics[width=0.92\columnwidth]{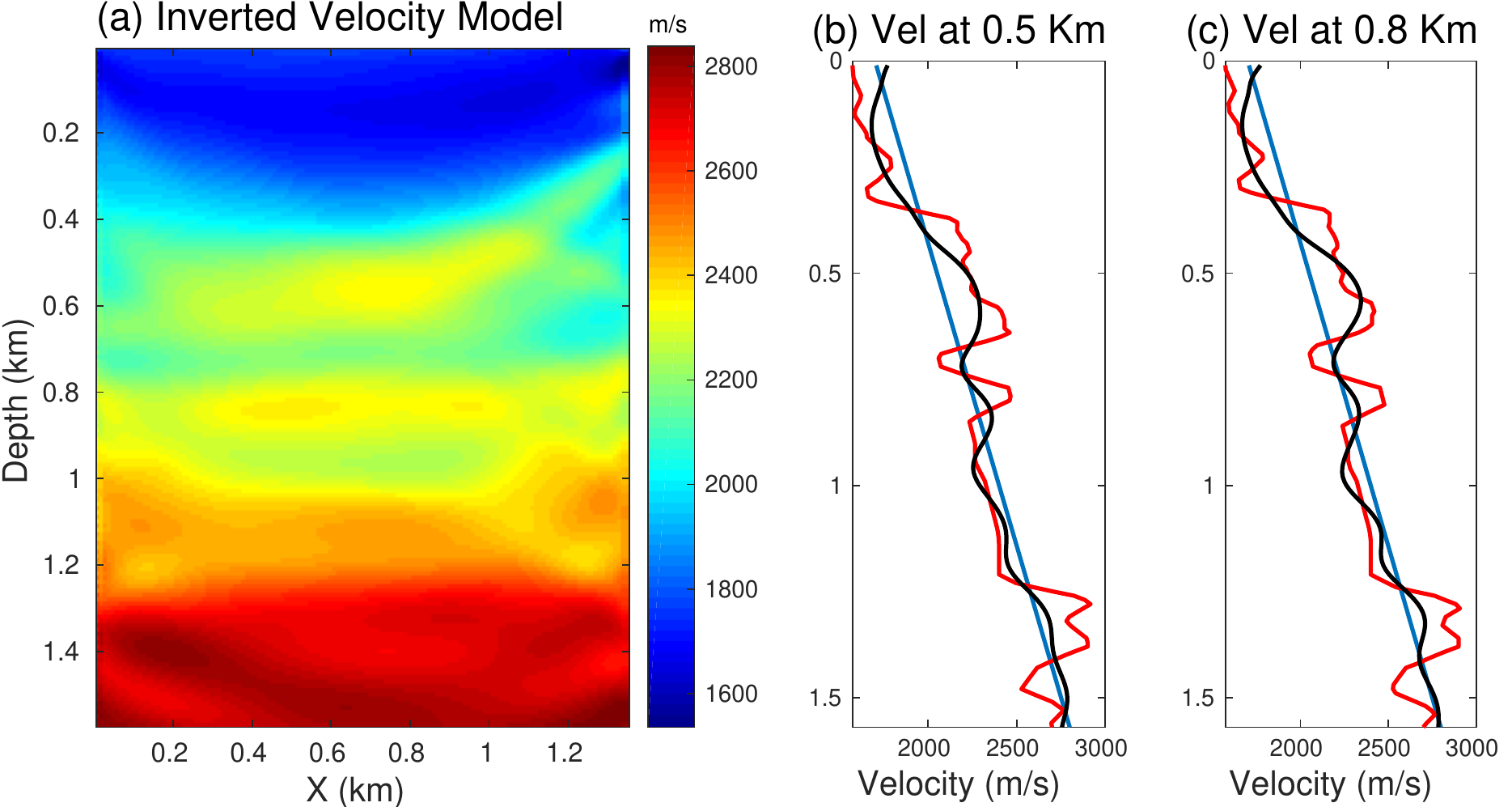}
\caption{The (a) inverted velocity model and the comparison of the vertical velocity profiles at (b) x = 0.5 km and x = 0.8 km. The blue, red and black curves indicate the velocity profiles of the initial, true and inverted velocity model, respectively.}
\label{fig:Mar2}
\end{figure}

Data computed from a part of the Marmousi model are used to test the skeletonized inversion method with the autoencoder method. We select the upper-right region of the Marmousi model shown in Figure \ref{fig:Mar1}a with 157 x 135 grid points. The finite-difference method is used to compute 77 acoustic shot gathers with 20 m source intervals along the depth of the well located at 10 m. Each shot contains 156 receivers that are evenly distributed at a spacing of 10 m along the vertical receiver well, which is located 1340 m away from the vertical source well. The data simulation time is 2 s with a time interval of 1 ms. The source wavelet is a 15 Hz Ricker wavelet and the initial model is shown in Figure \ref{fig:Mar1}b. Here we use the same autoencoder architecture and training strategy as was used in the previous numerical example. The inverted velocity model is shown in Figure \ref{fig:Mar2}a and the comparison of their vertical profiles at x = 0.5 and x = 0.8 are shown in Figure \ref{fig:Mar2}b and \ref{fig:Mar2}c, respectively. The blue, red and black curves represent the velocity profile of the initial, true and inverted velocity model, respectively. It shows that the inverted model is only able to reconstruct the low-wavenumber information in the true velocity model. To get a high-resolution inversion result, a hybrid approach such as the skeletonized inversion + full waveform inversion approach can be used \citep{luo1991wavea, luo1991waveb}. A plan for future research is to include a high-dimensional latent space.

\subsection{Friendswood Crosswell Field Data}
\begin{figure}[h]
\centering
\includegraphics[width=0.4\columnwidth]{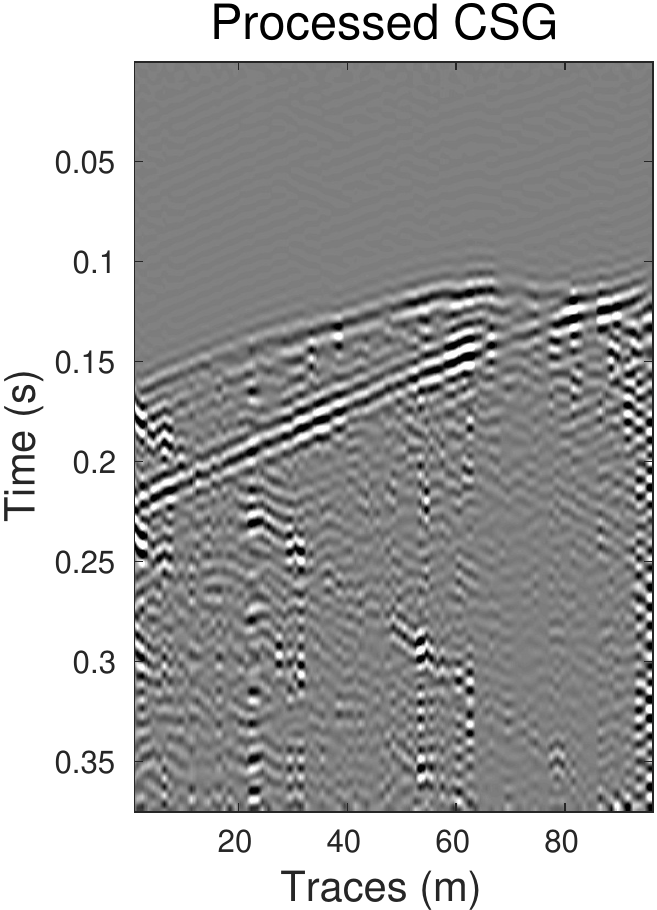}
\caption{A processed shot gather of Friendswoords data.}
\label{fig:Fri1}
\end{figure}

We now test our method on the Friendswood crosswell field data set. Two 305-m-deep cased wells separated by 183 m were used as the source and receiver wells. Downhole explosive charges were fired at intervals of 3 m from 9 m to 305 m in the source well, and the receiver well had 96 receivers placed at depths ranging from 3 m to 293 m. The data are low-pass filtered to 100 Hz with a peak frequency of 58 Hz. The seismic data were recorded with a sampling interval of 0.25 ms for total recording time of 0.375 s. However, we interpolate the data to 0.1 ms time interval for the numerical stable. A processed shot gather is shown in \ref{fig:Fri1}. Here, we mainly focus the inversion on the transmitted arrivals by windowing the input data around the early arrivals.

\begin{figure}[h]
\centering
\includegraphics[width=1.1\columnwidth]{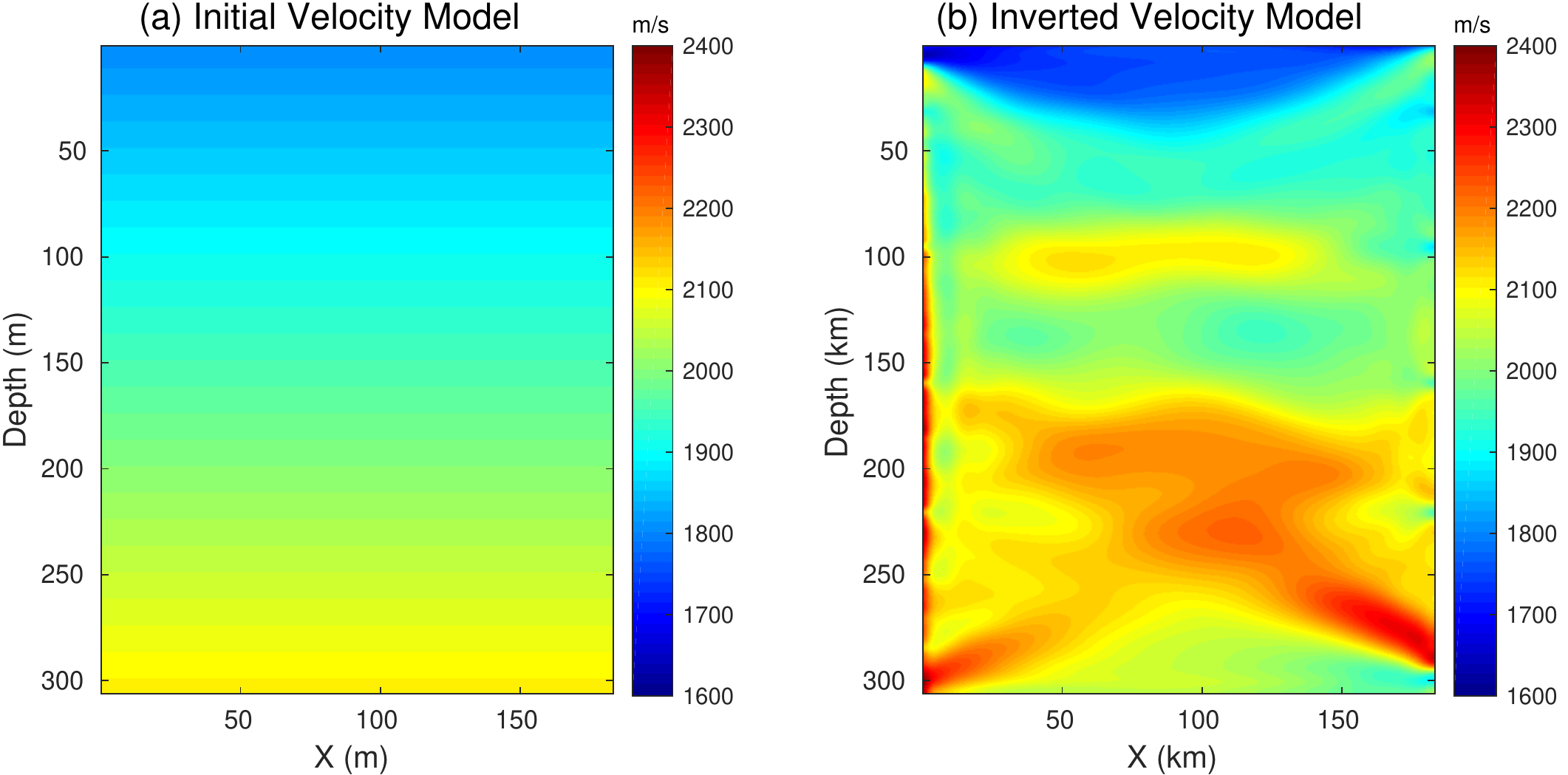}
\caption{The (a) initial linear increasing velocity and (b) inverted velocity models.}
\label{fig:Fri2}
\end{figure}

\begin{figure}[h]
\centering
\includegraphics[width=1.1\columnwidth]{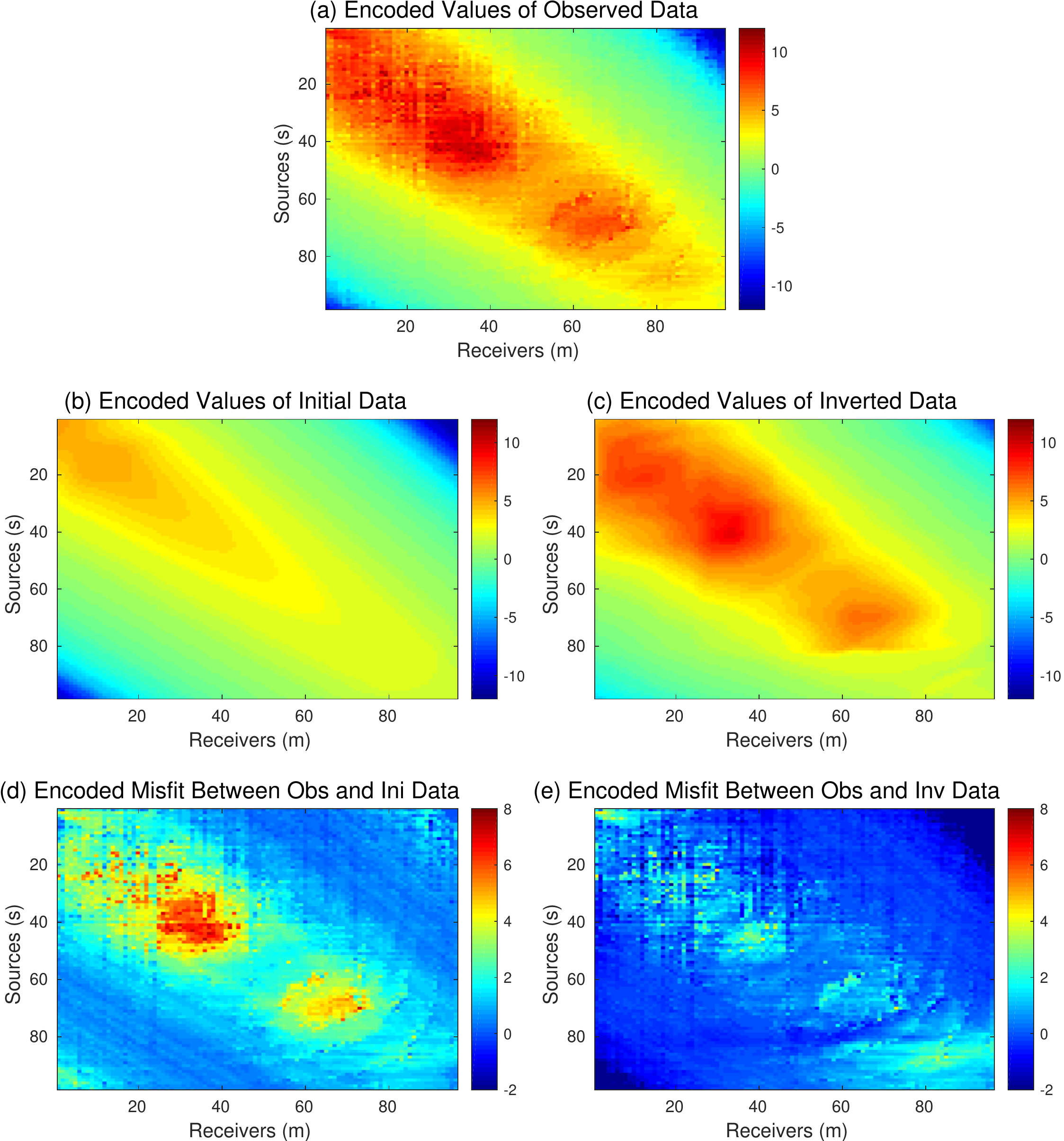}
\caption{The encoded value map of the (a) observed data, the synthetic data generated from the (b) initial model and (c) inverted model. The encoded misfit between the (d) observed data and initial data, (e) observed data and inverted data, respectively.}
\label{fig:Fri3}
\end{figure}

The autoencoder architecture we used here is almost the same as the previous two cases, except the dimensions of the input and output layer are changed to $3750 \times 1$. Only a portion of the observed data is used for training (every fifth shot gather is used for training). We do not stop the training until the misfit falls below a certain threshold. A linear increasing velocity model is used as the initial model which is shown in Figure \ref{fig:Fri2}a. Figure \ref{fig:Fri2}b shows the inverted velocity model with 15 iterations. Two high-velocity zones at the depth ranges between 85 to 115 m and 170 to 250 m appear in the inverted result. However, there are also some artifacts at the corners of the model that are due to statics and the geometry problems. Figure \ref{fig:Fri3}a shows the encoded value map of the observed data, where the vertical and horizontal axis represents the source and receivers indexes, respectively. It clearly shows that the near-offset traces have large positive values and the encoded values decrease as the offset increases. 

Figure \ref{fig:Fri3}b and \ref{fig:Fri3}c show the encoded value map of the seismic data generated from the initial and inverted velocity models, respectively, where the latter one is much more similar to the encoded value map of the observed data. To measure the distance between the true model and the initial model, we plot the values of the encoded misfit function in Figure \ref{fig:Fri3}d. It shows that there is a relatively larger misfit values at the near-offset traces than at the far offset traces. However, these misfits are largely reduced in the inverted tomogram that is shown in Figure \ref{fig:Fri3}e. This clearly demonstrates that our inverted tomogram is much closer to the true velocity model compared to the initial model.

\section{DISCUSSION}
Tests on both synthetic and observed data demonstrate that the wave equation inversion of seismic data skeletonized by an autoencoder can invert for the low-to-intermediate wavenumber details of the subsurface velocity model. To make this method practical we need to address the method's sensitivity to noisy data and wrapped effects.


\subsection{Noise Sensitivity Tests}
\begin{figure}[h]
\centering
\includegraphics[width=1.0\columnwidth]{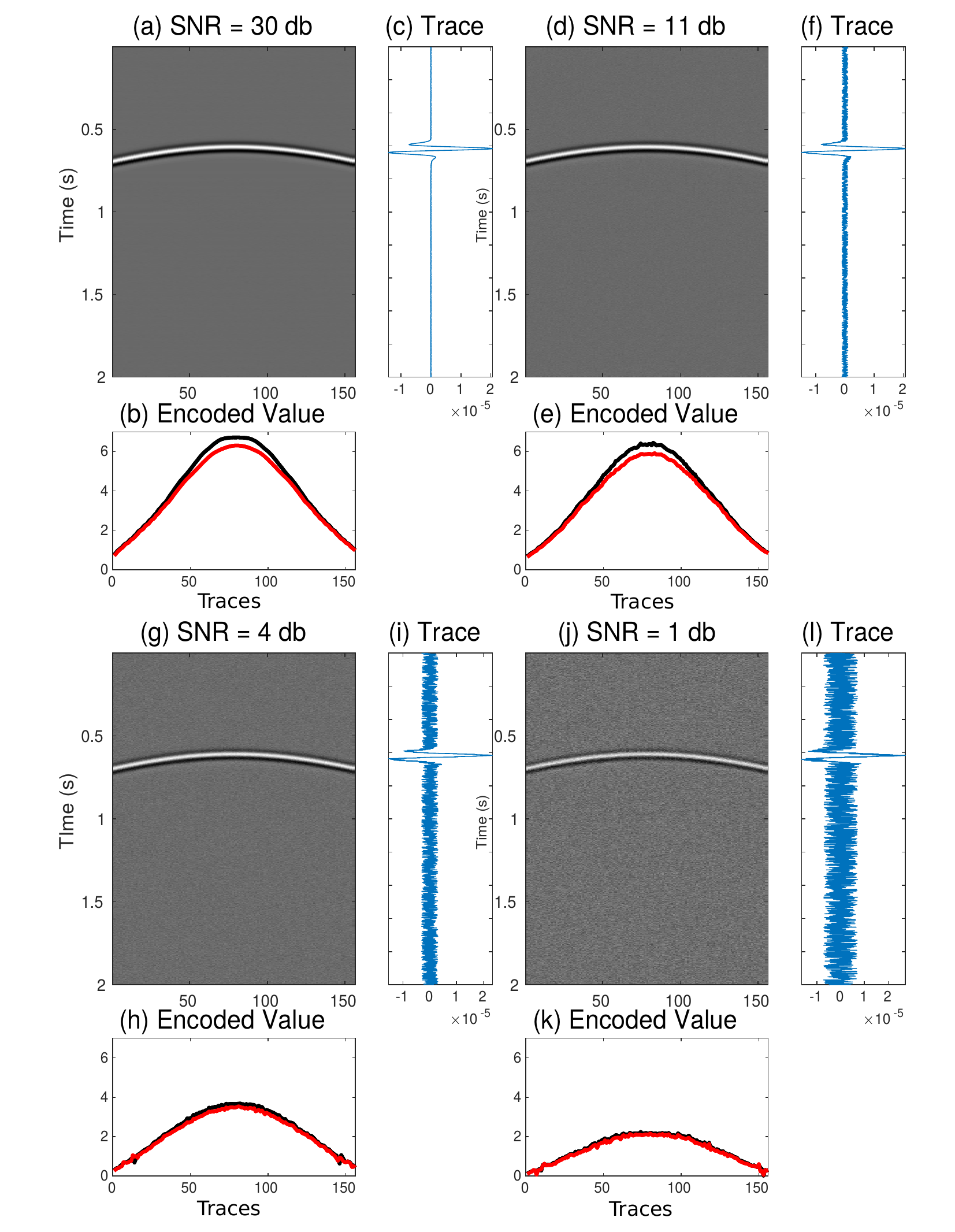}
\caption{The encoded value map of the (a) observed data, the synthetic data generated from the (b) initial model and (c) inverted model. The encoded misfit between the (d) observed data and initial data, (e) observed data and inverted data, respectively.}
\label{fig:diss1}
\end{figure}

In the previous synthetic tests we assumed that the seismic data is noise free. We now repeat the synthetic tests associated with Figure \ref{fig:Pro3}, except we add random noise to the input data. Different levels of noise are added on both the observed and synthetic data. Figure \ref{fig:diss1}a, \ref{fig:diss1}d, \ref{fig:diss1}g and \ref{fig:diss1}j show four shot gathers and their $80th$ traces are displayed in Figure \ref{fig:diss1}c, \ref{fig:diss1}f, \ref{fig:diss1}i and \ref{fig:diss1}l. Their encoded results are shown in Figure \ref{fig:diss1}b, \ref{fig:diss1}e, \ref{fig:diss1}h and \ref{fig:diss1}k, where the black and red curves represent the encoded values from the observed and synthetic data, respectively. It appears that the range of encoded values decreases as the noise level increases. Moreover, the encoded residual also decreases, which indicates that the encoded values becomes less sensitive to the velocity changes as the data noise level increase. 

\begin{figure}[h]
\centering
\includegraphics[width=1\columnwidth]{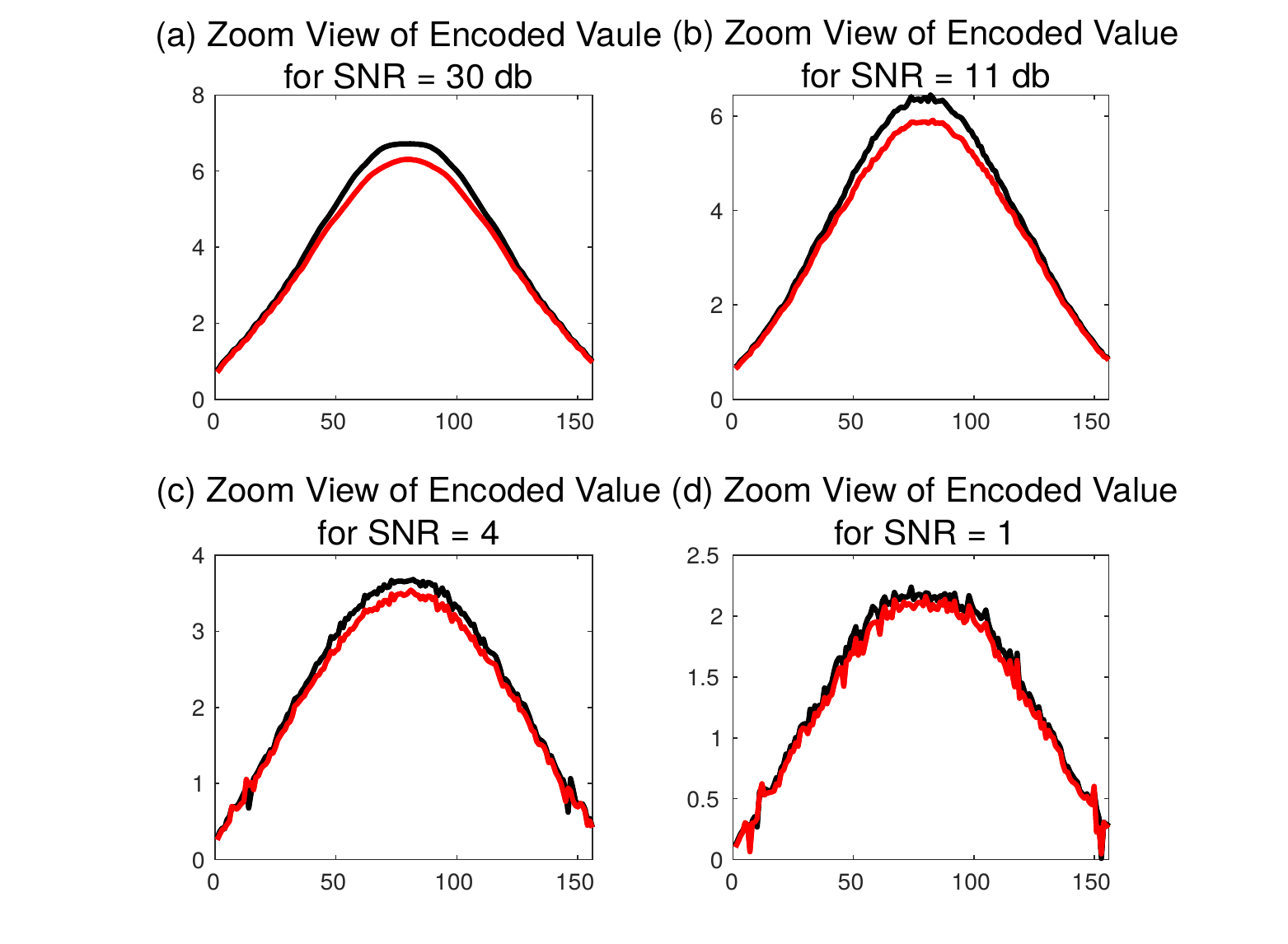}
\caption{The encoded value map of the (a) observed data, the synthetic data generated from the (b) initial model and (c) inverted model. The encoded misfit between the (d) observed data and initial data, (e) observed data and inverted data, respectively.}
\label{fig:diss2}
\end{figure}

Figure  \ref{fig:diss2} shows the zoomed views of the encoded values in Figure \ref{fig:diss1}, where some oscillations appear in the noisy data. These oscillations could further affect the accuracy of the inverted result, especially if the small velocity perturbation are omitted. Therefore, good data quality with less noise is preferred for the autoencoder method in order to recover an accurate subsurface velocity model.

\subsection{The Wrapped Effects of the Encoded Value}

\begin{figure}[h]
\centering
\includegraphics[width=1.1\columnwidth]{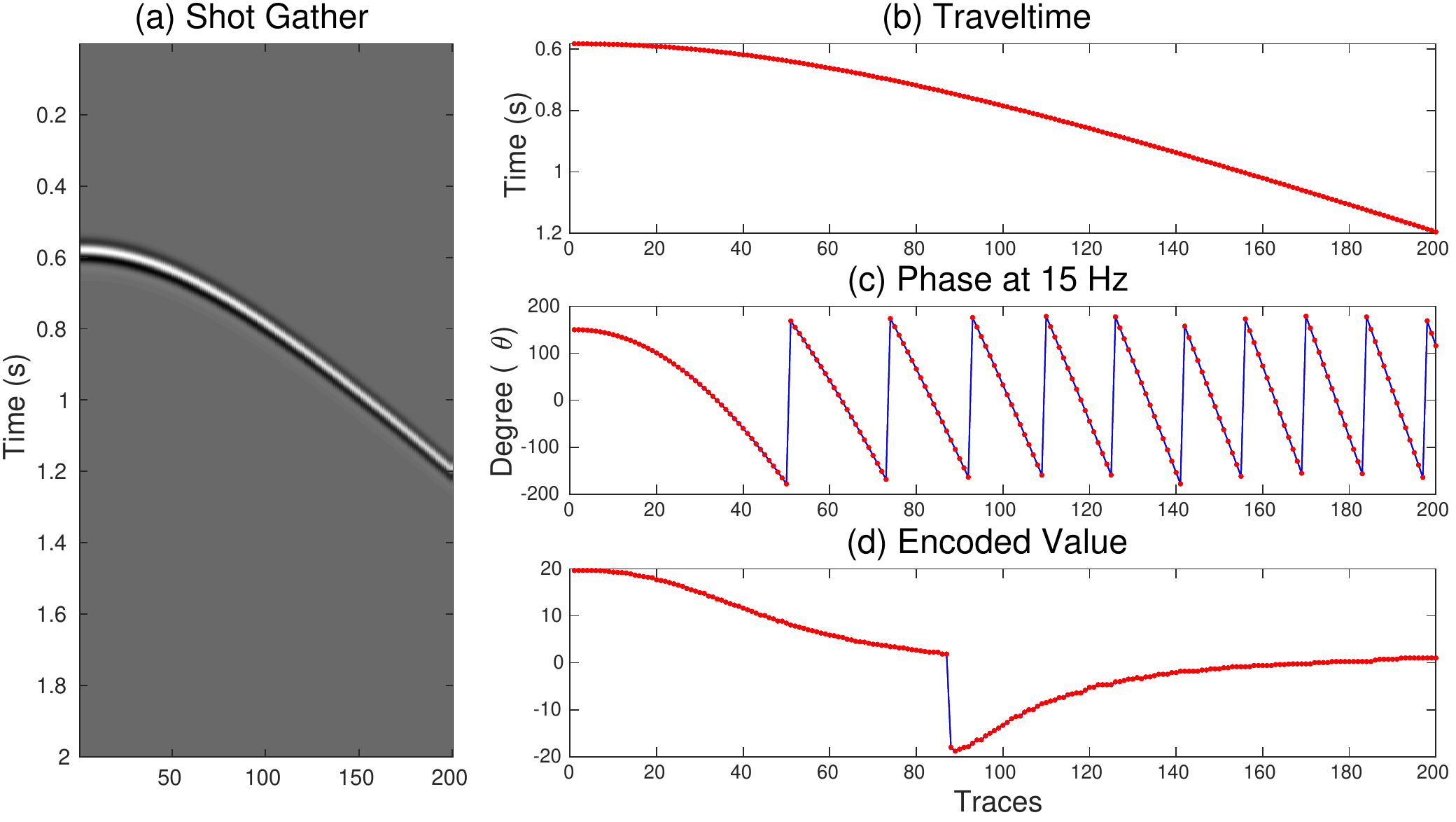}
\caption{The (a) shot gather and its corresponding (b) traveltime, (c) phases at 15 Hz and (d) encoded value.}
\label{fig:wrap1}
\end{figure}

The phase of a Fourier-transformed seismic signal has a finite but periodic range $[-\pi, \pi]$. Figure \ref{fig:wrap1}a shows a shot gather with an increasing traveltime delay along with the offset. Figure \ref{fig:wrap1}c shows that the phase changes for different traces at 15 Hz wraps around between $[-\pi, \pi]$, which is completely different with the traveltime changes (Figure \ref{fig:wrap1}b) that increase monotonically. Every time the phase reaches the $-\pi$ boundary, a sudden $2\pi$ jumps in phase happens. This wrap-around phenomenon is denoted as the wrapped effects of phase. The encoded value computed from the autoencoder also exhibits similar wrap effects when the traveltime variance of the seismic traces in the training set is large. Figure \ref{fig:wrap1}d shows the encoded results of the traces in Figure \ref{fig:wrap1}a, where a sudden jump occurs when the encoded values are close to zero. In this case, the encoded results should be unwrapped first before they are used for inversion. 


\section{CONCLUSIONS}
We introduce a wave equation method that finds the velocity model that minimizes the misfit function associated with the skeletonized data in the autoencoder's latent space. The autoencoder can compress a high-dimension seismic trace to a smaller dimension which best represents the original data in the latent space. In this case, measuring the encoded misift between the observed and synthetic data largely reduces the nonlinearity when compared with measuring their waveform differences. Therefore the inverted result will be less prone to getting stuck in a local minimum. The implicit function theorem is used to connect the perturbation of the encoded value with the velocity perturbation in order to calculate the gradient. Numerical results with both synthetic and field data demonstrate that skeletonized inversion with the autoencoder network can accurately estimate the background velocity model. The inverted result can be used as a good initial model for full waveform inversion.    

The most significant contribution of this paper is that it provides a general framework for using solutions to the governing PDE to invert skeletal data generated by any type of a neural network. The governing equation can be that for gravity, seismic waves, electromagnetic fields, and magnetic fields. The input data can be the records from different types of data, as long as the model parameters are sensitive to the model perturbations. The skeletal data can be the latent space variables of an autoencoder, a variational autoencoder, or feature map from a CNN, or PCA features. That is, we have combined the best features of Newtonian physics and the pattern matching capabilities of machine learning to invert seismic data by Newtonian machine learning.

\section{ACKNOWLEDGMENTS}
The research reported in this paper was supported by the King Abdullah University of Science and Technology (KAUST) in Thuwal, Saudi Arabia. We are grateful to the sponsors of the Center for Subsurface Imaging and Modeling (CSIM) Consortium for their financial support. For computer time, this research used the resources of the Supercomputing Laboratory at KAUST. We thank them for providing the computational resources required for carrying out this work. We also thank Schlumberger and BP for providing the BP2004Q data set and Exxon for the Friendswood crosswell data.

\bibliographystyle{seg}  
\bibliography{paper}

\newpage
\listoffigures

\end{document}